\documentclass{iau}
\usepackage{natbib}
\usepackage{graphicx}
\usepackage[usenames]{xcolor}

\usepackage[normalem]{ulem}

\def\d3k{{\displaystyle {\rm d}{\bf k} \over \displaystyle (2\pi)^3}}
\def\hmpc{h^{-1} {\rm Mpc}}

\title[Voids and the Cosmic Web]
{Voids and the Cosmic Web: \\ {\Large cosmic depression \& spatial complexity}}

\author[van de Weygaert]   
{Rien van de Weygaert$^1$}

\affiliation{$^1$Kapteyn Astronomical Institute, University of Groningen, \\ Postbus 800, NL-9700AD, Groningen, the Netherlands \\ email: weygaert@astro.rug.nl }

\pubyear{2014}
\volume{308}  
\pagerange{119--126}
\jname{The Zeldovich Universe: Genesis and Growth of the Cosmic Web}
\editors{R. van de Weygaert, S.F. Shandarin, E. Saar \& J. Einasto, eds.}

\begin{document}

\maketitle

\begin{abstract}
Voids form a prominent aspect of the Megaparsec distribution of galaxies and matter. Not only do they 
represent a key constituent of the Cosmic Web, they also are one of the cleanest probes 
and measures of global cosmological parameters. The shape and evolution of voids are highly sensitive to 
the nature of dark energy, while their substructure and galaxy population provides a direct key to the 
nature of dark matter. Also, the pristine environment of void interiors is an important testing ground 
for our understanding of environmental influences on galaxy formation and evolution. In this paper, we review 
the key aspects of the structure and dynamics of 
voids, with a particular focus on the hierarchical evolution of the void population. We demonstrate 
how the rich structural pattern of the Cosmic Web is related to the complex evolution and buildup 
of voids. 

\keywords{Cosmology, large-scale structure, voids, dark energy, modified gravity}
\end{abstract}

\firstsection 
\section{Introduction}
\begin{figure*}[t]
\begin{center}
\mbox{\hskip -0.18truecm\includegraphics[width=12.8cm]{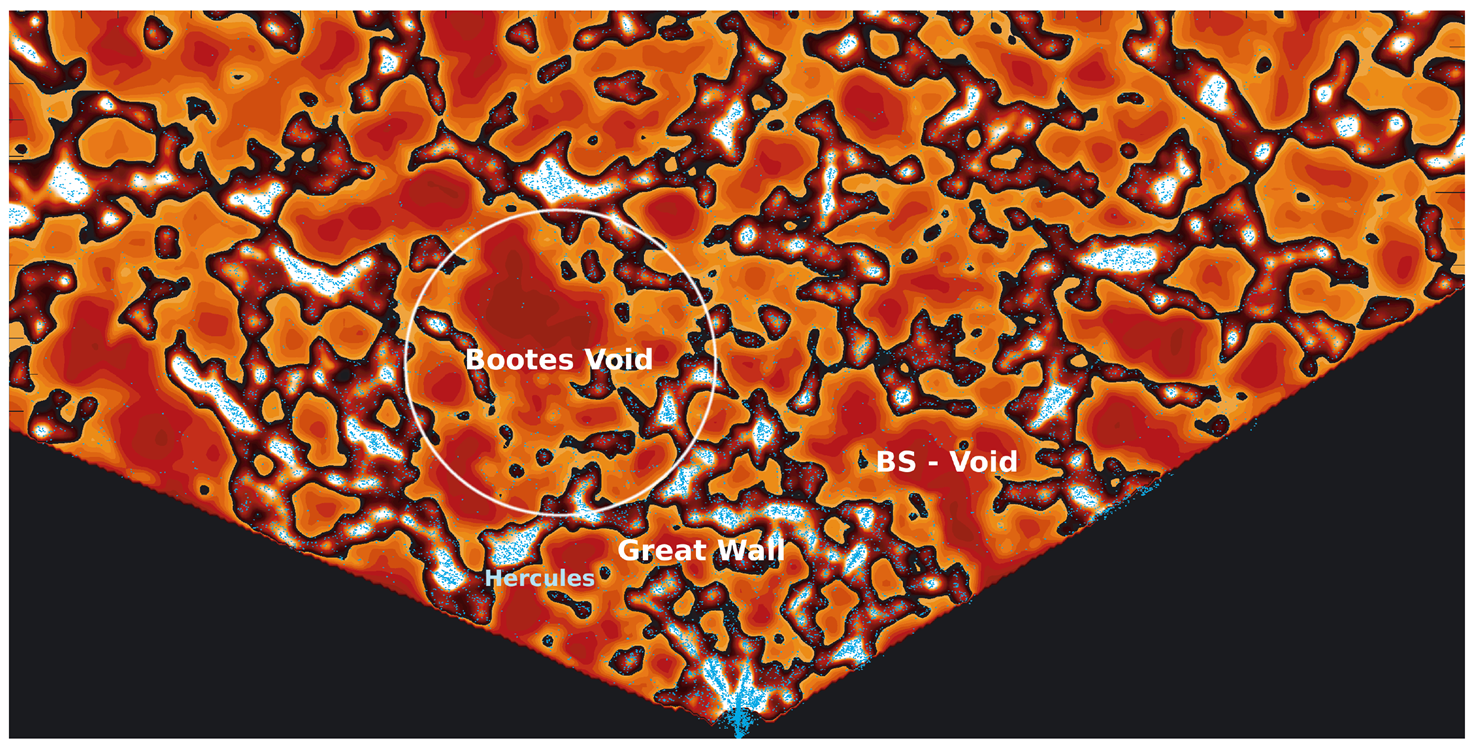}}
\end{center}
\begin{center}
\vskip 0.0truecm
\caption{SDSS density map and galaxies in the a region of the SDSS galaxy redshift survey region containing the 
canonical Bo\"otes void. The DTFE computed galaxy density map, Gaussian smoothed on a scale of $R_f=1 h^{-1} Mpc$, 
is represented by the color scale map. The galaxies in the SDSS survey are superimposed as dark dots. The underdense 
voids are clearly outlined as the lighter region outside the high-density weblike filamentary and wall-like features. 
We have indicated the location of the Hercules supercluster, the CfA Great Wall, and the Bo\"otes void.}
\end{center}
\label{fig:sdssvoids}
\end{figure*}

\noindent Voids form a prominent aspect of the Megaparsec distribution of galaxies and matter \citep{chincar1975,gregthomp1978,einasto1980,
kirshner1981,kirshner1987,lapparent1986,colless2003,tegmark2004,guzzo2013,guzzo2014}. They are enormous regions with sizes in the range
of $20-50h^{-1}$ Mpc that are practically devoid of any galaxy, usually roundish in shape and occupying the major share of space in the 
Universe (see fig.\ref{fig:sdssvoids} and \cite{weycospa2011} for a recent review).  Forming an essential and prominent aspect of the 
{\it Cosmic Web} \citep{bondweb1996}, they are instrumental in the spatial organization of the Cosmic Web 
\citep{icke1984,sahni1994,shethwey2004,aragon2010,einasto2011}. Surrounded by elongated filaments, sheetlike walls and 
dense compact clusters, they weave the salient weblike pattern of galaxies and matter pervading the observable Universe. 

Several recent studies came to the realization that voids not only represent a key constituent of the cosmic mass distribution, 
but that they are also one of the cleanest probes and measures of the global cosmology. Particularly interesting is the 
realization that their structure, morphology and dynamics reflects the nature of dark energy, dark matter and that of the 
possibly non-Gaussian nature of the primordial perturbation field. Another major aspect of voids is that their pristine environment 
represents an ideal and pure setting for the study of galaxy formation and the influence of cosmic environment 
on the evolution of galaxies. In addition, voids play a prominent role in the reionization process of the universe, forming the 
principal regions along which the ionizing radiation produced by the first stars in the Universe propages. 

In a void-based description of the evolution of the cosmic matter distribution, voids mark the transition scale at which 
density perturbations have decoupled from the Hubble flow and contracted into recognizable structural features. At any 
cosmic epoch the voids that dominate the spatial matter distribution are a manifestation of the cosmic structure 
formation process reaching a non-linear stage of evolution. On the 
basis of theoretical models of void formation one might infer that voids may act as the key organizing element for 
arranging matter concentrations into an all-pervasive cosmic network \citep{icke1984,regoes1991,weygaertphd1991,
shethwey2004,aragon2010b,aragon2013}. As voids expand, matter is squeezed in between them, and sheets and filaments form the void boundaries. 
This view is supported by numerical studies and computer simulations of the gravitational evolution of voids in 
more complex and realistic configurations \citep{martel1990,regoes1991,dubinski1993,weykamp1993,goldvog2004,colberg2005,
padilla2005,aragon2010b,aragon2013,sutter2014b,wojtak2016}. A marked example of the evolution of a typical large and deep void in a $\Lambda$CDM 
scenarios is given by the time sequence of six frames in fig.~\ref{fig:lcdmvoid}. 

The relatively simple structure and dynamical evolution of voids remains strongly influenced by the evolving large scale environment, 
through dominant tidal influences and direct contact between neighbouring voids. It is this aspect that has been 
recognized in a surge of recent studies for its considerable potential for measuring the value of cosmological parameters. The 
evolution of their structure and shape appear to be a direct manifestation of the nature of dark energy, while their dynamics reflects 
the nature of dark matter or that of the possibly nonstandard nature of gravity. 

\section{Depressions in the Universe}
\label{sec:webvoid}
Voids are an outstanding aspect of the weblike cosmic mass distribution \citep{lapparent1986,colless2003,huchra2012,pan2012,sutter2012}. 
They have been known as a feature of galaxy surveys since the first surveys were compiled \citep{chincar1975,
gregthomp1978,einasto1980,zeldovich1982}. Following the discovery by \cite{kirshner1981} and \cite{kirshner1987} of the most dramatic 
specimen, the Bo\"otes void, a hint of their central position within a weblike arrangement came with the first CfA redshift slice 
\citep{lapparent1986}. This view 
has been dramatically endorsed and expanded by the redshift maps of the 2dFGRS and SDSS surveys \citep{colless2003,tegmark2004}. 
They have established voids as an integral component of the Cosmic Web. The 2dFGRS maps and SDSS maps (see e.g. 
fig.~\ref{fig:sdssvoids}) are telling illustrations of the ubiquity and prominence of voids in the cosmic galaxy distribution. 
Recent studies have also revealed the prominence of voids in the distant and early U niverse. A beautiful example is that of the 
deep probe of the VIPERS galaxy redshift survey, which meticulously outlined the Cosmic Web up to redshifts $z\approx 0.5$ 
\citep{guzzo2013,guzzo2014}. Perhaps even more impressive is the identification of the considerably more intricate void population in 
the dark matter distribution, as revealed by its reconstruction in the 2MRS survey volume \citep{kitaura2012,hesskitaura2013} and the SDSS 
volume \citep{leclercq2015,leclercq2015b}. 
  
\begin{figure*}
\begin{center}
\mbox{\hskip -0.18truecm\includegraphics[width=10.0cm]{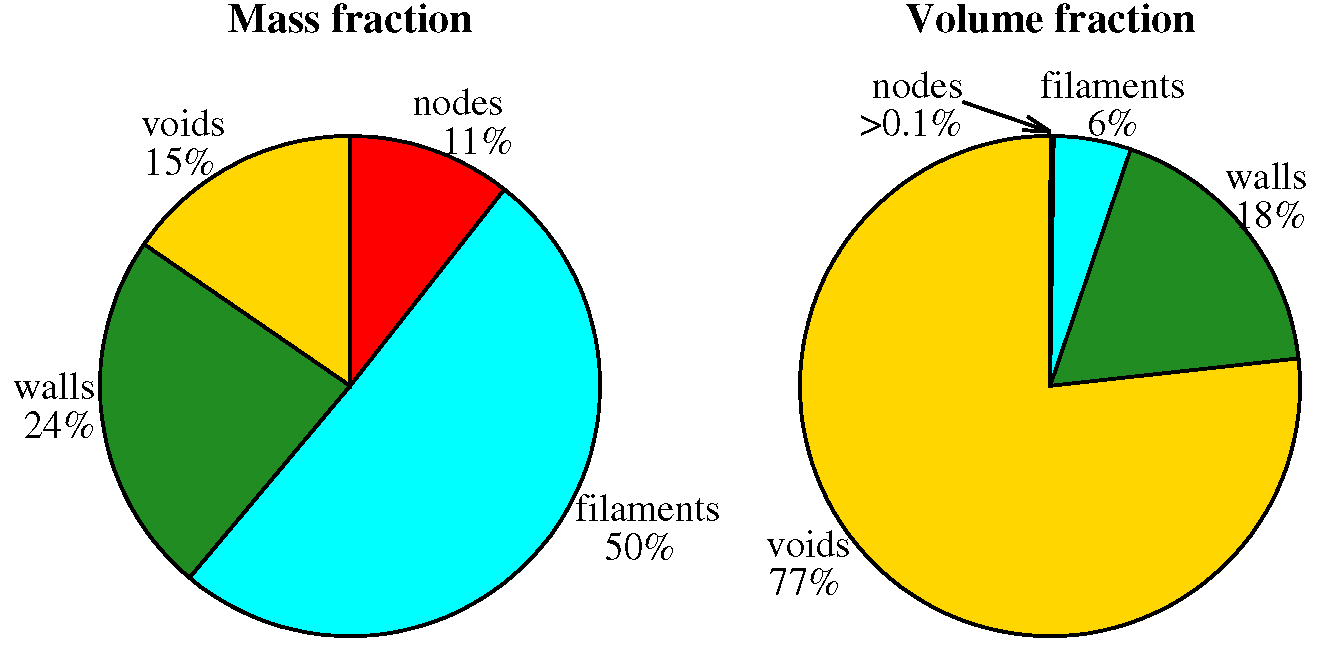}}
\end{center}
\begin{center}
\vskip 0.0truecm
\caption{The mass and volume fractions occupied by cosmic web environments detected by the NEXUS+ method. From Cautun et al. 2014}
\end{center}
\label{fig:voidinventory}
\end{figure*}

\bigskip
A relatively crude analysis of voids in computer simulations provides an impression of the status of voids in the Megaparsec 
universe. They clearly occupy a major share of the volume of the Universe. This has been confirmed in a recent 
systematic analysis by Cautun et al. (2014), which included an inventory of morphological components of the cosmic 
web with respect in the dark matter distribution in the LCDM Millennium simulation (see fig.~\ref{fig:voidinventory}). 
Around $77\%$ of the cosmic volume should be identified as a void region. Nonetheless, it represents less than $15\%$ of the mass 
content of the Universe. This implies that the average density in voids is $\approx 20\%$ of the average cosmic density, which is 
indeed reasonably close to what is expected on simple theoretical grounds (see below). 

\subsection{Voids and the Cosmos}
With voids being recognized as prominent aspects of the Megaparsec galaxy and matter distribution, we should 
expect them to be a rich source of information on a range of cosmological questions. We may identify at 
least four:
\begin{enumerate}
\item[$\bullet$] They are a prominent aspect of the Megaparsec Universe, instrumental in the spatial 
organization of the Cosmic Web \citep{ickewey1987,weykamp1993,shethwey2004,aragon2013}. Their effective repulsive influence 
over their surroundings has been recognized in surveys of the Local Universe \citep{courtois2012,tully2014}.
\item[$\bullet$] Voids contain a considerable amount of information on the underlying cosmological scenario and 
on global cosmological parameters. They are one of the cleanest cosmological probes and measures of dark energy, dark matter 
and tests with respect to possible modifications of gravity as described by General Relativity. This realization is based 
on the fact that it is relatively straightforward to relate their dynamics to the underlying cosmology, because they 
represent a relatively modest density perturbation. Their structure and shape, as well as mutual alignment, are direct reflections of dark energy 
\citep{parklee2007,leepark2009,platen2008,lavaux2010,lavaux2012,bos2012,pisani2015}. Notable cosmological imprints 
are also found in the outflow velocities and accompanying redshift distortions \citep{dekelrees1994,martel1990,rydmel1996}. 
In particular interesting is the realization that the dynamics and infrastructure of voids are manifestations of the 
nature of dark matter or of modified gravity \citep{li2011,peebles2010,clampitt2013,cai2015,cautun2016}.
The cosmological ramifications of the reality of a supersized voids akin to the identified ones by \cite{rudnick2007}, \cite{granett2009} 
and \cite{szapudi2015} would obviously be far-reaching.
\item[$\bullet$] The pristine low-density environment of voids represents an ideal and pure setting for the study 
of galaxy formation and the influence of cosmic environment on the formation of galaxies \citep[e.g.][]{kreckel2011,kreckel2012}.
Voids are in particular interesing following the observation by Peebles that the dearth of low luminosity objects in voids is hard to 
understand within the $\Lambda$CDM cosmology \citep{peebles2001}. 
\item[$\bullet$] Voids are prominent in the key reionization transition in the early universe,
key targets of LOFAR and SKA \citep[e.g.][]{furlanetto2006,morales2010}.
\end{enumerate}

\begin{figure*}
  \begin{center}
     \mbox{\hskip 0.0truecm\includegraphics[width=13.0cm]{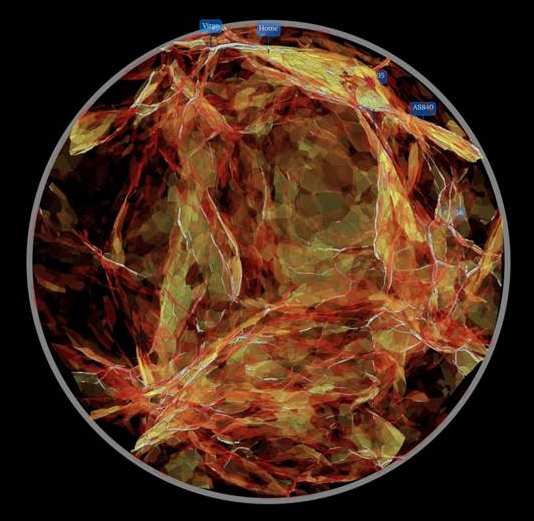}}
     \vskip 0.15cm
\caption{The Local Void. 3D rendering of the adhesion reconstruction of the Local Void. The reconstruction is based 
on the Bayesian KIGEN reconstruction by  Kitaura (2012) of the initial conditions in the local volume traced by 
the 2MRS redshift survey (He{\ss} et al. 2013). The Local Supercluster, including the Virgo cluster and the Local Group, are located at the 
top of the image.  Particularly striking is the precipitous emptiness of the Local Void. Image courtesy J. Hidding 
(see Hidding et al. 2016). }
\end{center}
\label{fig:localvoids}
\end{figure*}

\subsection{Voids in the Local Universe}
The most detailed view of the structure and galaxy population of voids is offered by voids in the local Universe. 
A particularly important source of information is the Karachentsev LV catalog of galaxies in the Local Volume, a 
volume-limited sample of galaxies within a radius of 11 Mpc around the Milky Way \citep{karachentsev2004}. It provides a meticulously 
detailed view of the Local Void that appears to dominate a major fraction of space in our immediate cosmic neighbourhood. The 
desolate emptiness of this vast volume is most strikingly borne out by the adhesion reconstruction of the Local Void by 
\cite{hidding2016} and \cite{hidding2016a}. 

\begin{figure*}
\vskip -0.25truecm
  \begin{center}
     \mbox{\hskip -0.55truecm\includegraphics[width=14.5cm]{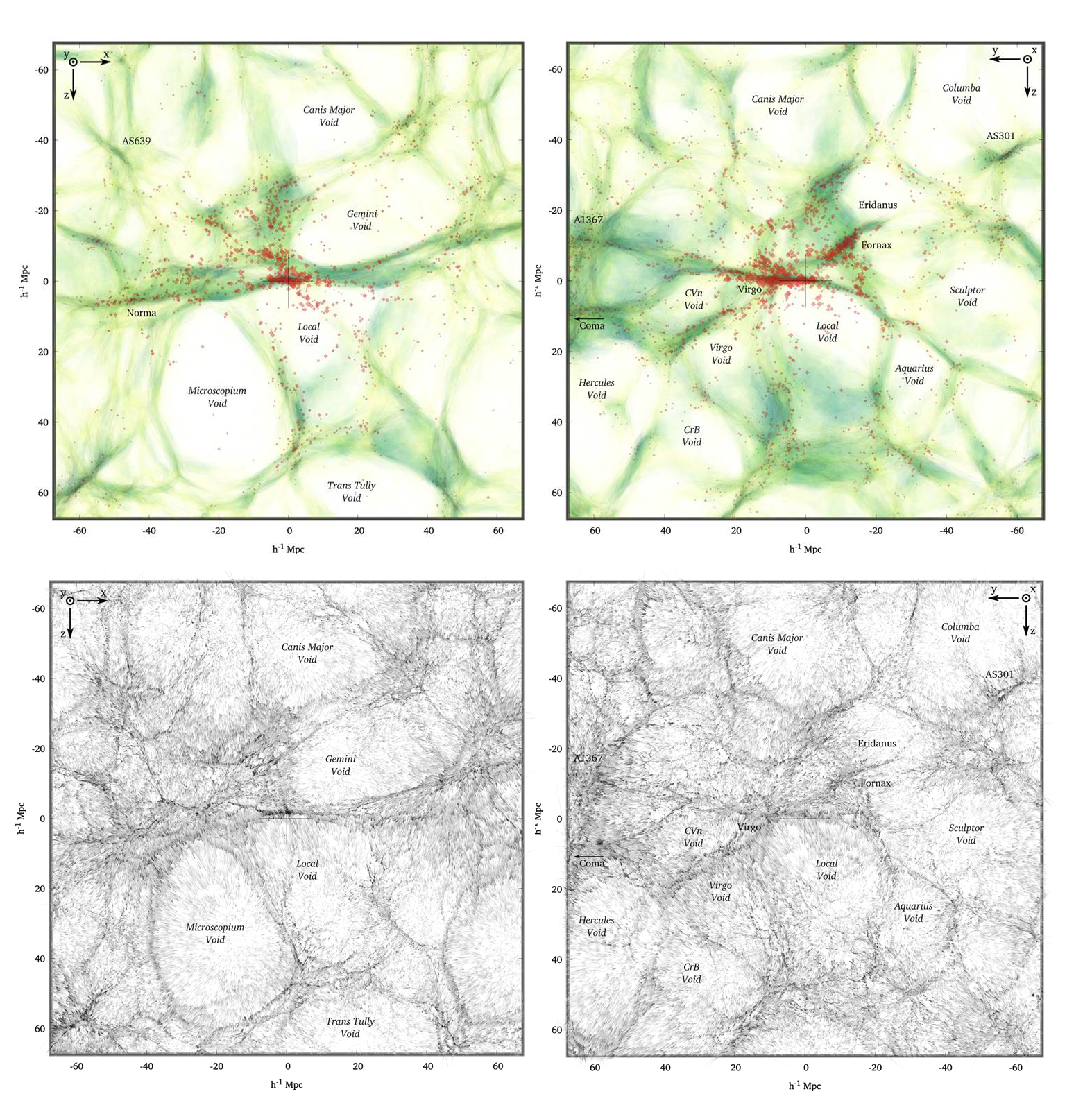}}
\caption{Voids in the Local Universe. The reconstruction of the weblike structure in the local Universe, 
sampled by the 2MRS survey, has been obtained on the basis of the adhesion formalism applied to a set 
of 25 constrained Bayesian KIGEN realizations of the primordial density and velocity field in the 
Local Universe (Kitaura 2012, He{\ss} et al. 2013). Top frames: (surface) density field along two perpendicular slices 
perpendicular to the plane of the Local Supercluster. Note that the density field concerns the dark matter distribution. The red dots 
are the 2MRS galaxies in the same volume. Bottom frame: the corresponding implied velocity flow in the same slices. 
Image courtesy J. Hidding (see Hidding et al. 2016). }
\end{center}
\label{fig:localvoids}
\end{figure*}

Moving to a slightly larger volume of the surrounding universe, we obtain a more representative impression of the 
prominence and role of voids in the overall large scale mass distribution. The 2MRS survey \citep{huchra2012} provides 
a uniquely complete census of the galaxy distribution out to distances of 100-150 Mpc. It entails the entire 
cosmic environment out to that distance, and has enabled remarkably precise and detailed reconstructions of the 
underlying matter distribution in our cosmic neighbourhood. 

Figure~\ref{fig:localvoids} provides a remarkably detailed reconstruction of the cosmic web in the 2MRS volume. It 
shows the (surface) density of the weblike structures in the Local Universe. These are the result of adhesion 
simulations by \cite{hidding2016}, based on the the constrained Bayesian KIGEN reconstruction by \cite{kitaura2012} 
of the initial conditions in the local volume traced by the 2MRS redshift survey \citep[see][]{hesskitaura2013}. For a given 
Gaussian primordial field, the adhesion formalism allows the accurate reconstruction of the rich pattern 
of weblike features that emerge in the same region as a result of gravitational evolution. The adhesion formalism 
was applied to 25 constrained realizations of the 2MRS based primordial density field \citep{hidding2012,hidding2016}. 
The mean of these realizations gives a reasonably accurate representation of the significant filamentary and wall-like 
features in the Local Universe. Most outstanding is the clear outline of the void population 
in the local Universe. The reconstruction also includes the velocity flow in the same cosmic region. It reveals the 
prominent nature of the outflow from the underdense voids, clearly forming a key aspect of the dynamics of the 
Megaparsec scale universe. 

The Local Universe structure in figure~\ref{fig:localvoids} presents a telling image of a void dominated 
large scale Universe. Many of the voids in the adhesion reconstruction can be identified with the void 
nomenclature proposed by Fairall \citep{fairall1998}, who mainly identified these voids by eye from the 
6dFGRS survey. It is interesting to see that the socalled Tully void appears to be a richly structured underdense 
region, containing at least the Microscopium Void, the Local Void and the "Trans Tully Void". 

\begin{figure*}
  \begin{center}
     \mbox{\hskip 0.0truecm\includegraphics[height=20.0cm]{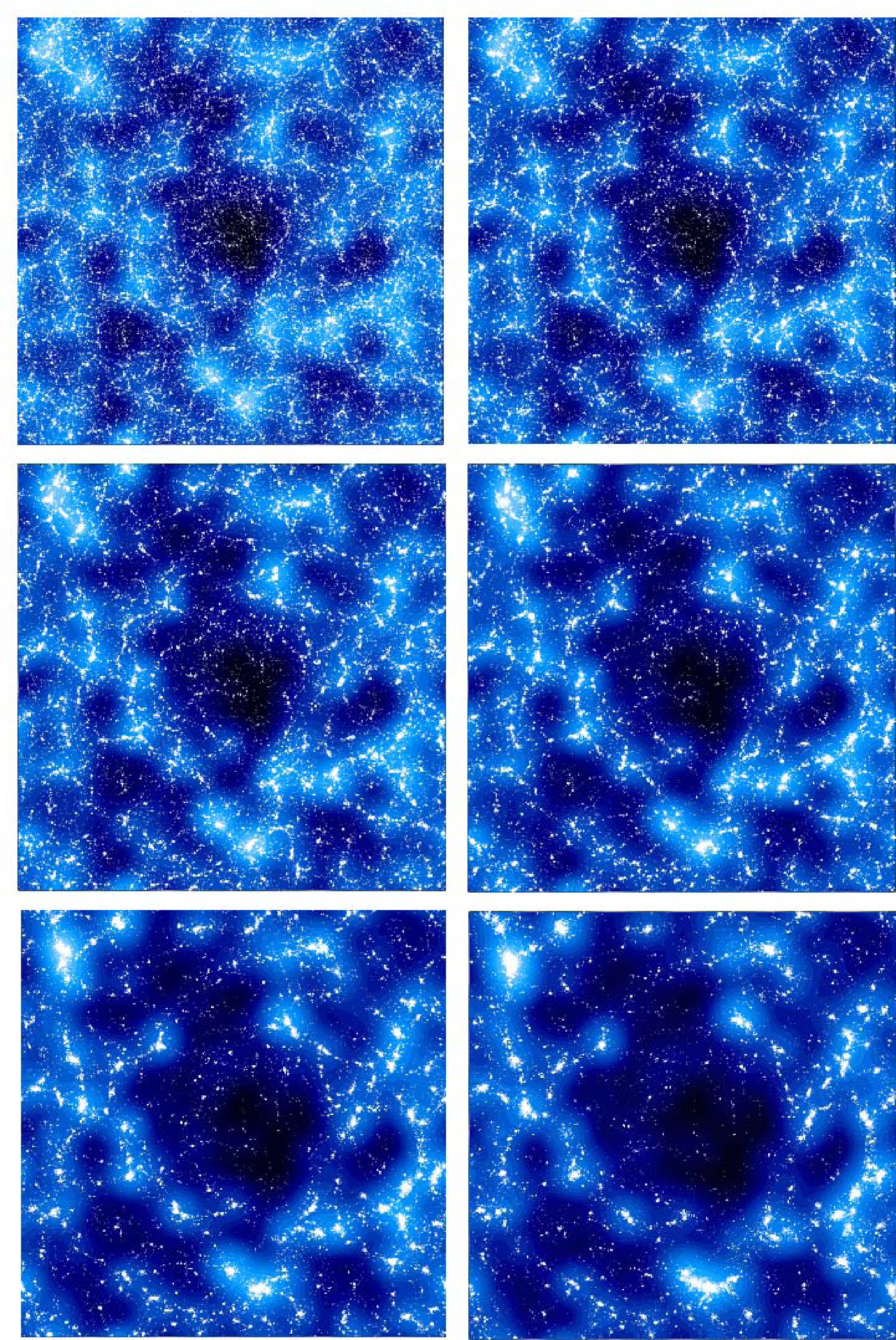}}
\caption{Simulation of evolving void (LCDM scenario). A void in a $n=0$ power-law power spectrum 
model. The slice is $50\hmpc$ wide and $10\hmpc$ thick. Shown are the particles and smoothed 
density field (smoothed on a scale of $4\hmpc$) at six different timesteps: a=0.05, 0.15, 0.35, 
0.55, 0.75 and 1.0. Image courtesy of Erwin Platen}
\end{center}
\label{fig:lcdmvoid}
\end{figure*}

\section{Formation and Evolution of Voids}
\label{sec:voidevol}
Voids emerge out of the density troughs in the primordial Gaussian field of density fluctuations. Early theoretical models 
of void formation concentrated on the evolution of isolated voids \citep{hoffshah1982,icke1984,edbert1985,blumenth1992}.  
Nearly without exception they were limited to spherically symmetric configurations \citep[but see][]{icke1984}. Neither of these 
simplifications appears to be close to what happens in reality. Nonetheless, the spherical model of isolated void evolution 
appears to provide us with valuable insights in major physical void characteristics and, as important, with sometimes 
surprisingly accurate quantitative reference benchmarks (see below). 

In the next section (sect.\ref{sec:ellipsoids}), we will see that a major and dominant aspect of voids is the fact that 
they are the opposite of isolated objects. To understand the dynamical evolution of voids we need to appreciate two major 
aspects in which 
the dynamics of voids differs fundamentally from that off dark halos. Because voids expand and increase in size, they 
will naturally run up against their expanding peers. Their spatial distribution and organization will be substantially 
influenced by the way in which the mutually competing voids distribute their share of space. Equally important is the 
fact that voids will always be limited to a rather modest density deficit: they cannot become more empty than empty.  
An immediate consequence of this is that the dynamical influence of the external mass inhomogeneities retains 
its dominant role in the evolution of a void. Voids will never and cannot decouple from their surroundings !

\subsection{Expansion, Evacuation, Dilution}
The essence of void evolution stems from the fact that they are underdensities in the mass distribution. As a result, they 
represent regions of weaker gravity. It translates in an effective repulsive peculiar gravitational influence. Most of 
the principal characteristics of void evolution can be recognized in the illustration of a typical 
evolving void in a $\Lambda$CDM Universe in fig.~\ref{fig:lcdmvoid} \citep[also see][]{weykamp1993}. The sequence of 
6 timestepe of the simulation reveals at least three of the principal aspects of void evolution. Because of the effective 
repulsive and thus outward peculiar gravitational accleration, initially underdense regions expand faster than the Hubble flow: 
they expand with respect to the background Universe. Also we see that mass is streaming out of the underdensity. As a result of this 
evacuation, the density within voids continuously decrease. A clear census of the continuously decreasing mass content in voids 
can be found in the study of cosmic web evolution by \cite{cautun2014}. Isolated voids would asympotically evolve towards an 
underdensity $\delta=-1$, pure emptiness. 

Also observable in figure~\ref{fig:lcdmvoid} is how the mass distribution inside the void appears to become increasingly 
uniform. At the same time we see an accumulation of mass around its boundary. This is a direct consequence of the differential outward 
peculiar gravitational acceleration in voids. Because the density within underdense regions gradually increases outward, we 
see a decrease of the corresponding peculiar (outward) gravitational acceleration. It means that matter at the centre of voids moves  
outward faster than matter at the boundary regions. As a result, matter tends to accumulate in - filamentary or planar - 
ridges surrounding the void, while the interior evolves into a uniform low-density region resembling a low-density homogeneous 
FRW universe \citep{goldvog2004}. 

Another key feature of void evolution is the diminishing prominence of substructure in its interior. Figure~\ref{fig:lcdmvoid} clearly 
shows how the internal structure gradually disappears as mass moves out of the void. This is a direct manifestation of the complex 
hierarchical evolution of voids \citep{shethwey2004}. A direct implication of this is that massive features and objects cannot form in 
voids, which e.g. manifests itself in a strong shift of the mass spectrum of dark halos - and hence galaxies - towards small 
masses \citep{cautun2014}. \cite{goldvog2004} showed how this can be rather accurately modelled by means of a modification of 
the spectrum of density fluctuations towards one more appropriate for a low-density FRW universe.

Note that while by definition voids correspond to density perturbations of at most unity, $|\delta_v|\leq 1$, mature voids 
in the nonlinear matter distribution do represent {\it highly nonlinear} features. This may be best understood 
within the context of Lagrangian perturbation theory \citep{sahnshan1996}. Overdense fluctuations may be described as 
a converging series of higher order perturbations. The equivalent perturbation series is less well behaved for voids:  
successive higher order terms of both density deficit and corresponding velocity divergence alternate between negative 
and positive.

\subsection{Spherical Voids}
\begin{figure*}
  \center
     \vskip -5,5truecm
     \mbox{\hskip -1.2truecm\includegraphics[width=16.0cm]{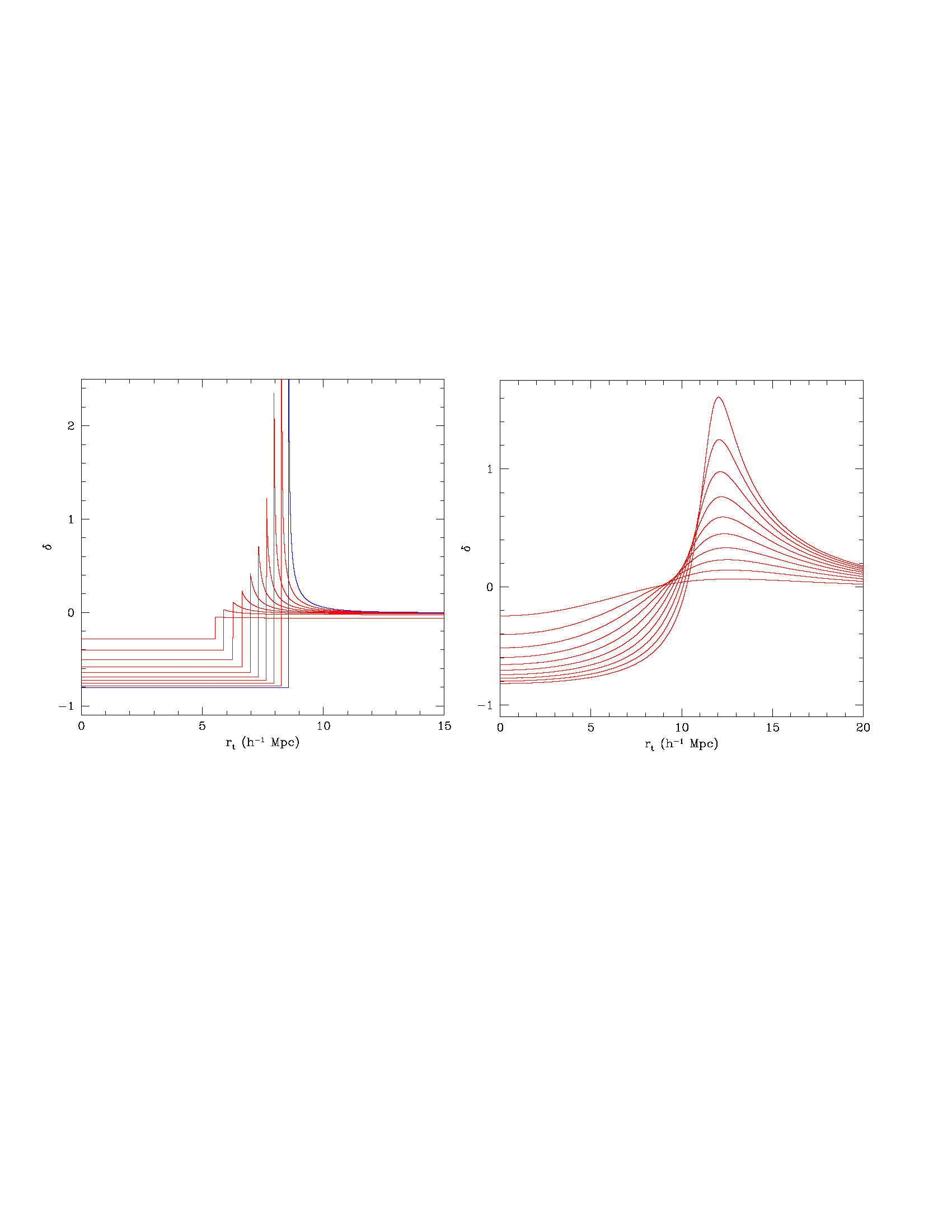}}
     \vskip -7.2truecm
\caption{Spherical model for the evolution of voids.  
 Left: a pure (uncompensated) "bucket" void evolving up to the epoch of 
 shell-crossing. Initial (linearly extrapolated) density deficit was 
 $\Delta_{lin,0}=-10.0$, the initial (comoving) radius ${\widetilde R}_{i,0}=5.0 h^{-1}\hbox{Mpc}$.  
Righthand: a void with an angular averaged 
SCDM profile. Initial density deficit 
and characteristic radius are same as for  the tophat void (left). The 
tendency of this void to evolve into a tophat configuration by the time of 
shell crossing is clear. Shell-crossing, and the formation of a ridge, 
 happens only if the initial profile is sufficiently steep.}
\label{fig:sphervoid}
\end{figure*}

While in reality voids will never be isolated, nor spherical, we may obtain a basic understanding of the 
quantitative aspects of void evolution from the simple model of an evolving isolated spherical void. 
Figure~\ref{fig:sphervoid} illustrates the evolution of spherical isolated voids. The void in the lefthand 
is a typical pure and uncompensated ``bucket" void\footnote{This is more commonly known as {\it tophat}, but given the 
configuration, {\it bucket} seems a more appropriate description.}. We notice the principal characteristics of void 
formation that we discussed in the previous section. The expansion of the void is evident as we see the boundary edge, and the 
ridge at its boundary, move outward. Meanwhile, matter moves out of the void, leading to an increasingly empty 
void. 

The spherical void model also clarifies an additional major aspect, the formation of a ridge around the 
expanding void. This is the result of the differential outward expansion of the mass in and around the void. 
While the uniform underdensity profile implies a uniform expansion within the void, near the boundary the 
inner layers move outward faster than the more outward layers. This is the result of the latter feeling a 
more moderate interior underdensity. In all, this ultimately leads to the interior mass shells taking over 
the initially exterior shells. This leads to a {\it fundamental} evolutionary timescale for void evolution, that 
of {\it shellcrossing}. 

\cite{edbert1985} demonstrated that once voids have passed the stage of shellcrossing they enter a phase of self-similar expansion. 
Subsequently, their expansion will slow down with respect to the earlier linear expansion. This impelled \cite{blumenth1992} to 
identify voids in the present-day galaxy distribution with voids that have just reached the stage of shell-crossing. It happens 
when a primordial density depression attains a linearly extrapolated underdensity $\delta_v=f_v=-2.81$ (strictly speaking for an EdS universe). That 
happens when a perfectly spherical "bucket" void will have expanded by a factor of 1.72 at shellcrossing, and therefore have evolved into an underdensity 
of $\sim 20\%$ of the global cosmological density, ie. $\delta_{v,nl}=-0.8$. In other words, the voids that we see nowadays in the galaxy 
distribution do probably correspond to regions whose density is $\sim 20\%$ of the mean cosmic density (note that it may be different 
for underdensity in the galaxy distribution). 

Interestingly, for a wide range of initial radial profiles, voids will evolve into a bucket shaped 
void profile. The righthand frame of fig.\ref{fig:sphervoid} shows the evolution of a spherical 
void that develops out of an initial underensity that is an angular averaged density profile of an 
underdensity in a CDM Gaussian random field \citep[see][]{bbks,weyedb1996}. Due to the differential 
expansion of the interior mass shells, we get an acccumulation of mass near the exterior and boundary 
of the void, meanwhile evening out the density distribution in the interior. We also recognized this 
behaviour in the more complex circumstances of the LCDM void in fig.~\ref{fig:lcdmvoid}. To 
a considerable extent this is determined by the steepness of the density profile of the protovoid 
 depression \citep{palmvogl1983}. In nearly all conceivable situations the void therefore appears to 
assume a {\it bucket shape}, with a uniform interior density depression and a steep outer boundary 
(fig.~\ref{fig:sphervoid}, righthand frame). The development of a {\it bucket} shaped density profile 
may therefore be considered a generic property of cosmic voids.

Recently, there have been a range of studies on the issue of void density profiles, and the question 
whether they display universal behaviour \citep[see e.g.][]{hamaus2014,cautun2016}. In a range of 
studies, Hamaus et al. \citep[e.g][]{hamaus2014} concluded that spherically averaged density profiles 
of voids do indeed imply a universal density profile, that could be parameterized by 2 factors. 
Interestingly, these density profiles have a less prominent bucket shaped interior profile than 
those seen for the spherical voids in fig.~\ref{fig:sphervoid}. This may be understood from the 
fact that voids in general are not spherical (see sec.~\ref{sec:voidshapes}), so that spherical averaging 
will lead to the mixing of different layers in the void's interior. The recent study by \cite{cautun2016} 
confirms this: when taking into account the shape of voids, a remarkably strong bucket 
void density profile appears to surface. 

The corresponding void expansion velocity profile confirms the above. A uniform density distribution 
within a void's interior directly translates into a superHubble velocity outflow, ie. an outflow 
in which the expansion velocity scales linear with the distance to the void's center. In the next 
section we will discuss this within the context of the dynamics of voids. 

\section{Void Dynamics}
\begin{figure*}[t]
  \centering
    \mbox{\hskip -0.5truecm\includegraphics[width=9.5cm,angle=270.0]{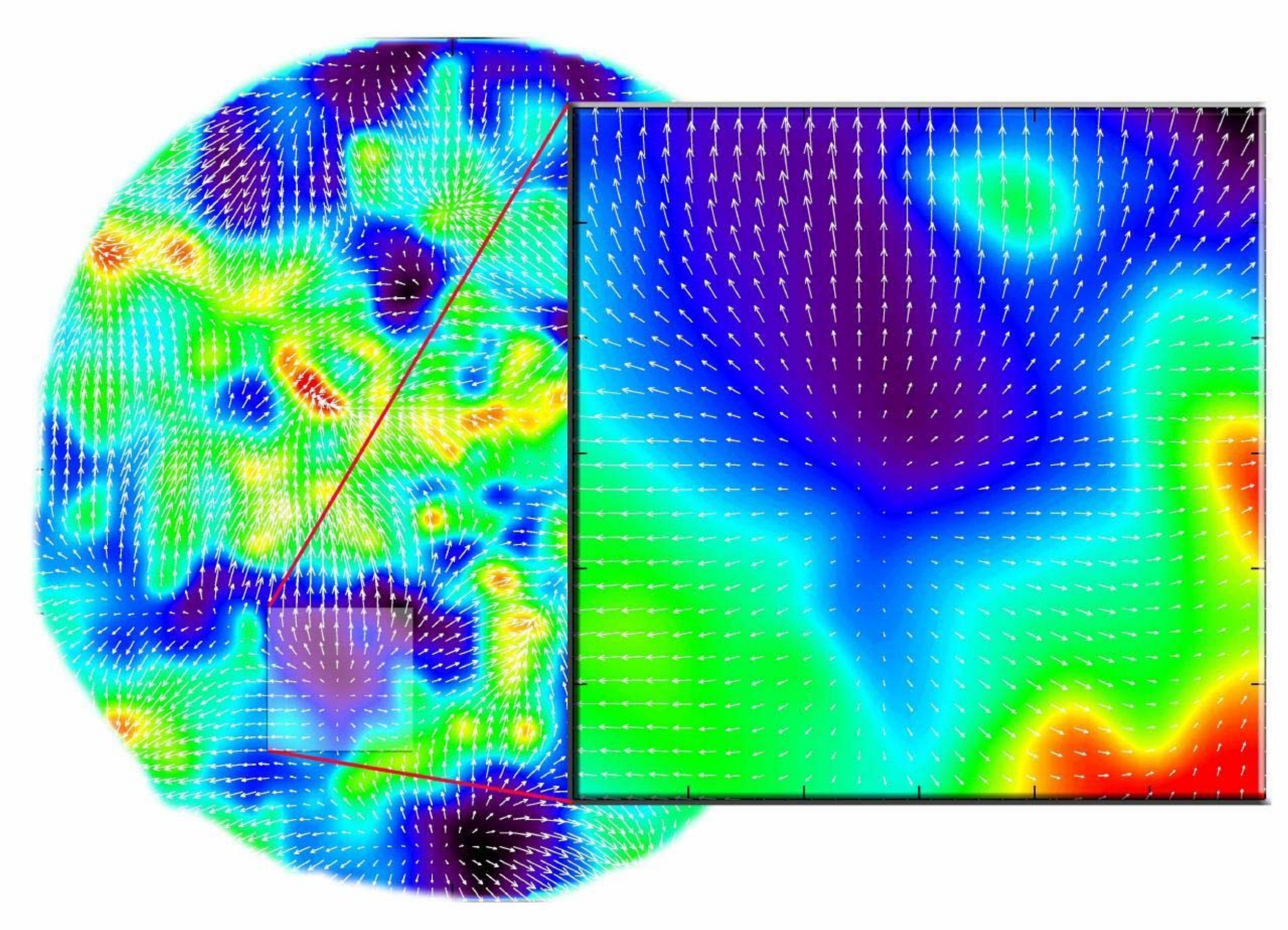}}
    \vskip -0.0truecm
    \caption{Gravitational impact of the Sculptor Void. The righthand frame shows the inferred velocity field 
    in and around the Sculptor void near the Local Supercluster. The colour map represents the density values, 
    with dark blue at $\delta \sim -0.75$ and cyan near $\delta \sim 0.0$. The vectors show the 
    implied velocity flow around the void, with a distinct nearly spherically symmetric outflow. 
    It is a zoom-in onto the indicated region in the density and velocity map in the Local Universe (lefthand) 
    determined on the basis of the PSCz galaxy redshift survey. From: Romano-D\'{\i}az \& van de Weygaert  2007.
} 
\label{fig:psczsculptor} 
\end{figure*}

Soon after their discovery, various studies pointed out their essential role in the organization of the cosmic matter 
distribution \citep{icke1984,regoes1991}, based on the realization that they have a substantial dynamical influence on 
their surroundings. Voids have an effective repulsive influence over their surroundings (see below). Even though they represent 
relatively limited density deficits - voids will never exceed $|\delta|>1$ - the fact that they occupy a major share 
of cosmic volume translates into a dominant factor in the dynamical interplay of forces on Megaparsec scales. The flow in and 
around the void is dominated by the outflow of matter from the void, culminating into the void's own expansion near the outer 
edge. This has indeed been recognized in various galaxy surveys and surveys of galaxy peculiar motions in our Local Universe. 
A telling illustration is the map of velocity flows implied by the PSCz galaxy redshift survey (see fig.~\ref{fig:psczsculptor}). The 
gravitational impact of the Sculptor Void on our immediate cosmic vicinity is directly visible. The influence of voids 
on the peculiar velocities of galaxies have been even recognized at an individual level \citep{tully2008}. 

To fully to understand the dynamical influences on the evolution of voids, and as well to 
appreciate the dynamical impact of voids on their environment, we may obtain substantial insight from the 
idealized configurations of a spherically isolated void, and that of the force field on 
a homogeneous ellipsoidal void. The spherical model allows us to understand the principal 
aspects of the flow field in the interior of voids. The ellipsoidal model allows us to 
evaluate the importance of external forces on the dynamical evolution of voids. 

In terms of their dynamics, the key observation is that external mass inhomogeneities retain
a dominant role in the evolution of void. Key to this realization is the fact that voids are 
always be limited to a rather modest density deficit of not more than $\delta=-1$. This  will 
strongly limit their own local gravitational action. It also means that the implicit tidal influences 
induces by the large scale environment of voids represent a major factor in their dynamical evolution. 
Their shape, mutual alignment and overall expansion are aspects that are strongly dependent on 
these external tidal influences. 

It also means that global cosmological influences retain a strong impact on the void's 
evolution. While highly nonlinear overdensities internally hardly notice the presence of the repulsive 
force induced by the presence of dark energy, in the desolate interiors of voids dark 
energy plays an even  considerably more prominent role than on global cosmological scales 
\citep{parklee2007,leepark2009,lavaux2010,lavaux2012}. Along the same line, it is in the diluted density 
field of a void that the 
imprint of possible modifications of the force of gravity will be most noticeable \citep{li2011,clampitt2013}. 
While this direct dynamical influence of dark energy, and gravity modifications, in voids may be one factor 
of importance, there is an equally and possibly even more sizeable secondary influence. The dominant role of 
external tidal influences on the void's dynamics depends on the amplitude of the inhomogeneities in the 
large scale environment of voids. The growth rate of mass fluctuations is a sensitive function of factors 
such as the nature of dark energy, and the nature of gravity on large scales, as these influence the 
dynamical timescales involved in the formation of structure. 

To fully appreciate the forces at work in voids, we first looks at their 
own principal imprint, that of the void's expansion and the corresponding outflow 
velocities. Subsequently, we will investigate the external influences on the basis of 
the ellipsoidal model.

\subsection{Superhubble Bubbles}
Because voids are emptier than the rest of the universe, they have an outward directed 
velocity flow: mass is flowing out of the void. Evidently, the resulting void velocity 
field profile is intimately coupled to that of its interior density field profile.  

In the situation of a mature, evolved void, the velocity field of a void resembles that 
of a Hubble flow, in which the outflow velocity increases linearly with distance to the 
void center. In other words, voids are {\it super-Hubble bubbles} \citep{icke1984} ! 
The linear velocity increase is a manifestation of its general expansion. The near constant 
velocity divergence within the void conforms to the {\it super-Hubble flow} expected for the near 
uniform interior density distribution that voids attain at more advanced stages (see fig.\ref{fig:sphervoid}).

It is straightforward to appeciate this from the {\it continuity equation}. For a uniform 
density field, it tells us that the velocity divergence in the void will be uniform, 
corresponding to a Hubble-like outflow. Because voids are emptier than the rest of 
the universe they will expand faster than the rest of the universe, with a net 
velocity divergence equal to
\begin{eqnarray}
   \theta&\,=\,&{\displaystyle \nabla\cdot{\bf v} \over \displaystyle H}\,=\,3 (\alpha-1)\,,\qquad\alpha=H_{\rm void}/H\,,
\end{eqnarray}
\noindent where $\alpha$ is defined to be the ratio of the super-Hubble expansion rate of the 
void and the Hubble expansion of the universe. Inspection of the velocity flow in the Sculptor 
void (see fig.\ref{fig:psczsculptor}) shows that this is indeed a good description of the observed 
reality. 

\begin{figure*}[t]
  \centering
    \mbox{\hskip -5.5truecm\includegraphics[width=9.0cm]{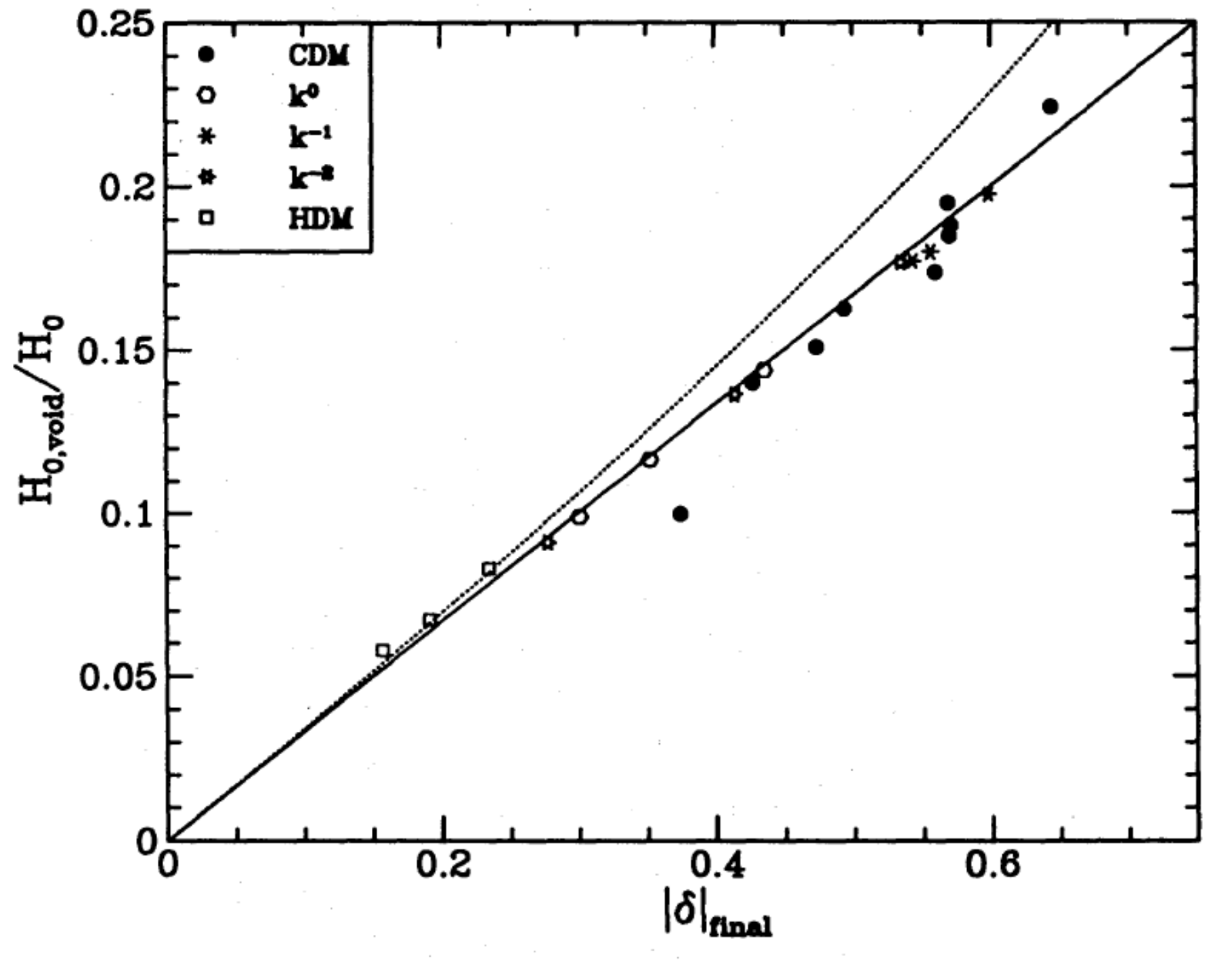}}
    \vskip 1.0truecm
    \mbox{\hskip -1.0truecm\includegraphics[width=13.5cm]{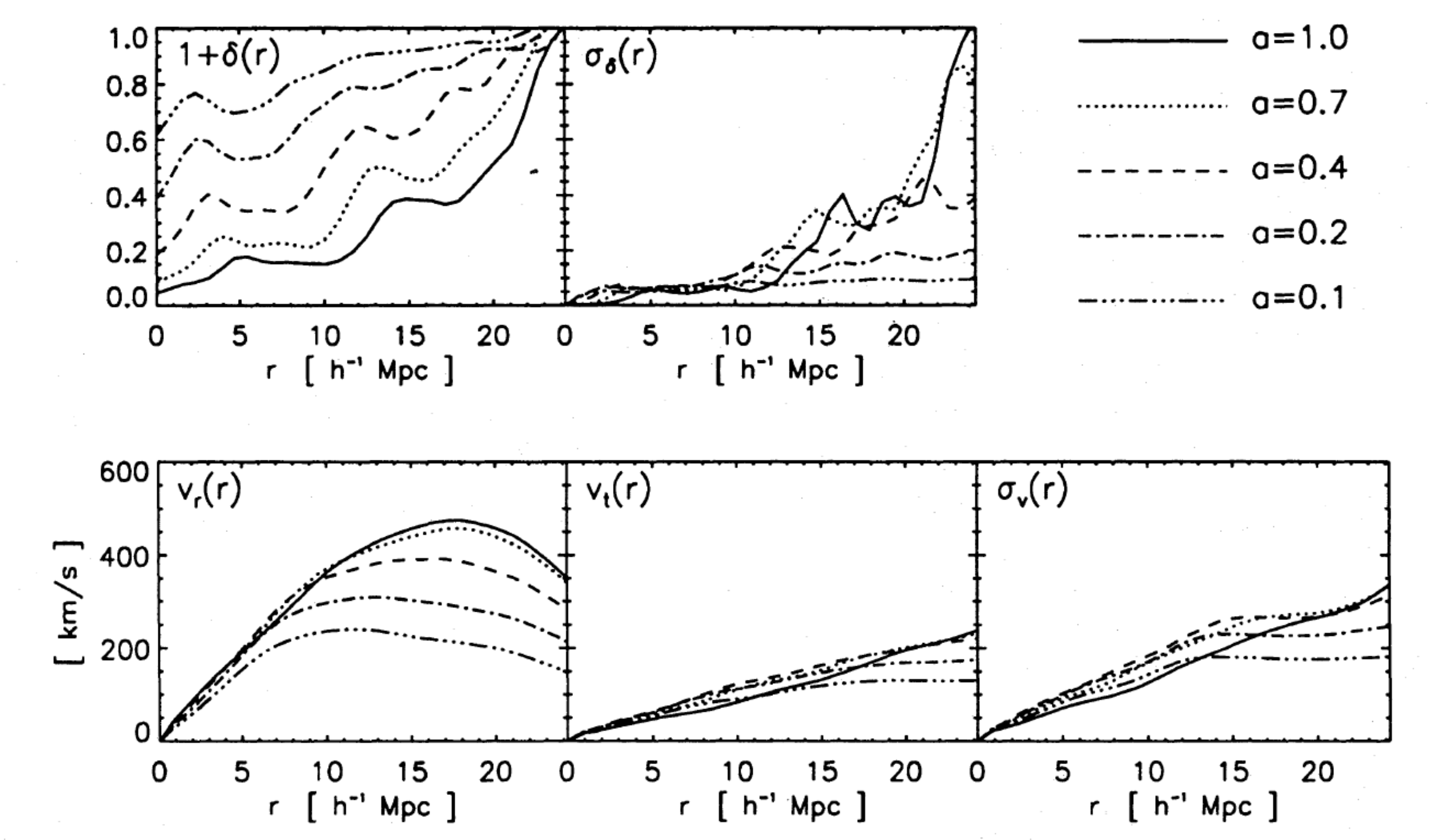}}
    \vskip -0.0truecm
    \caption{Superhubble Void expansion. Bottom rows: the density profiles and velocity 
profiles of an evolving void in a constrained simulation of a void in a CDM cosmology. 
Clearly visible is the evolution of the void towards a bucket shaped density profile, 
and the corresponding development of a linear superHubble flow field in the interior of 
the void.  Top: the superHubble void expansion rate, 
in units of the global Hubble parameter $H(t)$ for a set of simulated voids in various 
cosmologies, as a function of the underdensity $\Delta(t)$ of a void (here called $\delta_{final}$. 
From: van de Weygaert \& van Kampen 1993. 
} 
\label{fig:superhubble} 
\end{figure*}

\cite{weykamp1993} confirmed that the velocity outflow field in viable cosmological scenarios 
does indeed resemble that of a superHubble expanding bubble (see fig.~\ref{fig:superhubble}b). 
They also established that the superHubble expansion rate is directly proportional to the nonlinear 
void density $\Delta(t)$ (see fig.~\ref{fig:superhubble}a), 
\begin{equation}
H_{\rm void}/H\,=\,-{\displaystyle 1 \over \displaystyle 3} f(\Omega)\,\Delta(t)\,.
\end{equation}
This relation, known within the context of a linearly evolving spherical density perturbation, 
in the case of fully evolved voids appears to be valid on the basis of the {\it nonlinear} 
void density deficit. As \cite{weykamp1993} illustrated, voids should therefore be considered 
as distinctly nonlinear objects. Several recent studies \citep[e.g.][]{hamaus2014} have 
confirmed this finding for voids in a range of high resolution cosmological simulations. 

\subsection{Tidal Forces - Ellipsoidal Voids}
\label{sec:ellipsoids}
Of decisive importance for understanding the dynamics and evolution of voids is the realization 
that voids will never and cannot decouple from their surroundings. One aspect of this is that 
voids expand and increase in size, they will naturally run up against their expanding peers. Their 
spatial distribution and organization will be substantially influenced by the way in which the mutually 
competing voids distribute their share of space. 

The second strong environmental influence on the evolution of voids is that of the tidal 
influences induced by the large scale environment of voids. Their shape, mutual alignment and 
overall expansion are aspects that are strongly dependent on these external tidal influences. 

\subsubsection{Homogeneous Ellipsoidal Model}
Arguably one of the most direct and transparent means of developing our insight into the tidally 
affected dynamics of voids is to look at the evolution of homogeneous underdense ellipsoids. This idea was forwarded 
by \cite{icke1984}. In particular the dominant role of external tidal forces on the evolution of 
voids can be understood by assessing the evolution of homogeneous underdense ellipsoids, within 
the context of {\it Homogeneous Ellipsoidal Model} \citep{icke1973,whitesilk1979,eisenstloeb1995,bondmyers1996,
desjacques2008}. 

It is interesting to realize that the ellipsoidal model, even while idealized, is a rather good 
approximation for the main aspects of a void's evolution. In many respects the homogeneous model is far 
more suitable as an approximation for underdense regions than it is for overdense ones: voids expand and 
evolve towards an increasingly uniform interior density field over a vast range of their volume (see 
fig.~\ref{fig:sphervoid}): while they expand their interior gets drained of matter and 
develops a flat ``bucket-shaped'' density profile. Overdense regions contract into more 
compact and hence steeper density peaks, so that the area in which the ellipsoidal model represents 
a reasonable approximation will continuously shrink. Evidently, the approximation is restricted to the 
interior and fails at the void's outer fringes. The latter is a result of the accumulation of material 
near the void's edge, and the encounter with neighouring features of the cosmic web and surrounding voids. 

\bigskip
The homogeneous ellipsoidal model assumes an object to be a region with a triaxially symmetric 
ellipsoidal geometry and a homogeneous interior density, embedded within a uniform background 
density $\rho_{\rm u}$. Consider the simple situation of the external tidal shear 
directed along the principal axes of the ellipsoid. The gravitational acceleration along 
the principal axes of an ellipsoid with over/underdensity $\delta$ can be evaluated from the 
expression for the corresponding scale factors ${\cal R}_{\rm i}$,
\begin{eqnarray}
{\displaystyle d^2 {\cal R}_{\rm m} \over \displaystyle d t^2}&\,=\,&- 4 \pi G \rho_{\rm u}(t)\, 
\left[{1+\delta\over 3}\,+\,{1\over 2}\,(\alpha_m-{2 \over 3})\,\delta\right]\,{\cal R}_{\rm m}\,-\,\tau_{\rm m}\,
{\cal R}_{\rm m}\,+\,\Lambda R_m\,,
\label{eq:ellipseqnmot}
\end{eqnarray}
where we have also taken into account the influence of the cosmological constant $\Lambda$. The factors 
$\alpha_{\rm m}(t)$ are the ellipsoidal coefficients specified by the integral equation,  
\begin{eqnarray}
\alpha_{\rm m}(t)\,=\,{\cal R}_1(t) {\cal R}_2(t) {\cal R}_3(t)\,{\int}^\infty_0\,
{\displaystyle {\rm d}\lambda\ \over \displaystyle ({\cal R}_{\rm m}^2(t)+\lambda )\,
\prod\nolimits_{n=1}^3 \left({\cal R}_{\rm n}^2(t)+\lambda\right)^{1/2}}\,.
\label{eq:ellalph}
\end{eqnarray}

\bigskip
\noindent The influence of the external (large-scale) tidal shear tensor $T^{(ext)}_{\rm mn}$ enters via the 
eigenvalue $\tau_m$. From eqn.~\ref{eq:ellipseqnmot}, it is straightforward to appreciate that 
as $\delta$ grows strongly nonlinear, the relative influence of the large-scale (near-)linear 
tidal field will decline. However, the density deficit $\delta$ of voids will never exceed 
unity, $|\delta|<1$, so that the importance of the factor $\tau_m$ remains relatively large. 

\begin{figure*}[t]
\begin{center}
\vskip -8.0truecm
\mbox{\hskip -0.25truecm\includegraphics[width=15.0cm]{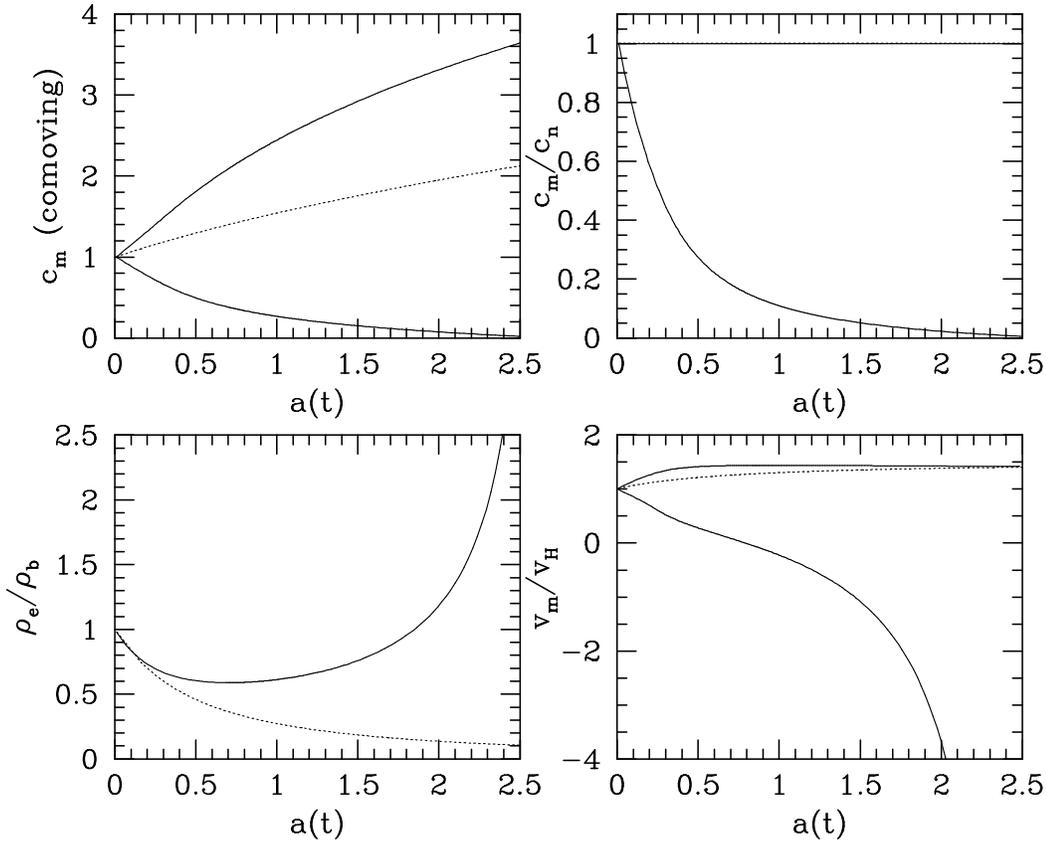}}
\vskip -0.25truecm
\caption{Anisotropic Void Expansion and Collapse. Evolution of the 
same initially spherical void. Dashed line: without external influences  
Solid line: under the influence of an axisymmetric and linearly evolving 
external tidal force field ($E_{mm,0}=(-E,-E,2E)$).
{\it Topleft:} (comoving) void axis, $c_m'=c_m/a(t)$. 
{\it Topright:} axis ratios $c_2/c_1$ and $c_3/c_1$. 
{\it Bottom Left:} Internal density $\rho_e$ of the void, in 
units of the cosmic density $\rho_u(t)$. {\it Bottom Right:} 
The velocity $v_m$ along the axes of the voids, in units of the 
Hubble velocity $v_H$. Note the collapse along axis 1 and 2.}
\label{fig:voidcoll}
\end{center}
\end{figure*}

The impact of the external tidal forces can be so strong they not only effect an anisotropic expansion of the 
void, but even may manage to make an initially spherically expanding void to collapse. 
The latter can be seen in fig.~\ref{fig:voidcoll}, where we compare the evolution of the 
two initially spherical and equivalent void regions. The isolated one assumes the regular evolution 
of a spherical isolated void, while its peer develops into an increasingly anisotropic configuration. 
At some point the external tides go as far to squeeze the void into contraction and ultimately towards 
collapse. This is the situation that we find in the hierarchical buildup of voids, to be described 
in the next section~\ref{sec:voidsocio}. As illustrated in the accompanying illustration of the 
hierarchically evolving void population, small voids near the boundary of large void bubbles get 
squeezed into collapse as a result of the tidal impact of their overdense surroundings.  

\subsubsection{Void Shapes and Tides}
\label{sec:voidshapes}
\cite{icke1984} pointed out that any (isolated) aspherical underdensity will become more spherical as it expands. 
The effective gravitational acceleration is stronger along the short axis than along the longer axes. For 
overdensities this results in a stronger inward acceleration and infall, producing increasingly flattened 
and elongated features. By contrast, for voids this translates into a larger {\it outward} acceleration 
along the shortest axis so that asphericities will tend to diminish. For the interior of voids this 
tendency has been confirmed by {\rm N}-body simulations \citep{weykamp1993}. 

In reality, voids will never reach sphericity. Even though voids tend to be less flattened or elongated than 
the halos in the dark matter distribution, they are nevertheless quite nonspherical: \cite{platen2008} find 
that they are slightly prolate with axis ratios of the order of $c:b:a\approx 0.5:0.7:1$. This agrees 
with the statement by \cite{shandarin2006} and \cite{parklee2007} that in realistic cosmological circumstances 
voids will be nonspherical. This is also quite apparent in images of simulations, such as those in 
figure~\ref{fig:voidhieradh}.

The flattening is a result of large scale dynamical and environmental factors, amongst which we can identify at 
least two important factors \citep{platen2008}. Even while their internal dynamics pushes them to a more 
spherical shape they will never be able to reach perfect sphericity before encountering surrounding structures 
such as overdense filaments or planar walls. Even more important is the fact that, for voids, the external tidal
influences remain important (see discussion above). These external tidal forces are responsible for a significant 
anisotropic effect in the development of the voids. In extreme cases they may even induce fullscale collapse 
and demolition of the void.   

\subsubsection{Void Shape Measurement: the WVF Watershed Void Finder}
Figure~\ref{fig:devoids} illustrates a typical example of the shape distribution of voids in a $\Lambda$CDM 
cosmology. The identification of voids by means of the watershed transform, which delineates the region of a 
void independent of its scale and shape, a direct objective measurement of the volume, shape and orientation of the void 
population is obtained. This may be directly inferred from the comparison between the bottom panels with the 
dark matter distribution in the top panel of fig.\ref{fig:devoids}. Introduced and proposed by \cite{platen2007} (also 
see contribution by Jones \& van de Weygaert in this volume), 
the Watershed Void Finder (WVF) identifies voids via a watershed transform applied to the DTFE (Delaunay Tessellation 
Field Estimator) density field reconstruction \citep{schaapwey2000,weyschaap2009}. The latter exploits the scale 
and shape sensitivity of Voronoi and Delaunay tessellations to retain the intricate multiscale and weblike nature of the 
mass distribution probed by a discrete particle or galaxy distribution. The idea of the use of the watershed 
transform for the objective identification of voids, given the close relation to the topology of the 
weblike Megaparsec matter distribution \citep[see][]{aragon2010}, has in the meantime been recognized as 
a true watershed with respect to setting a standard definition for what should be considered as 
a void \citep{colberg2008,neyrinck2008,vide2015,nadathur2015}.

\subsubsection{Void Alignments}
\label{sec:voidalign}
Large scale tidal influences not only manifest themselves in the shaping of individual voids. They are 
also responsible for a distinct alignment of substructures along a preferred direction, while they 
are also instrumental in their mutual arrangement and organization. Locally, the orientation of a void 
turns out to be strongly aligned with the tidal force field generated by structures on scales up to at 
least $20-30h^{-1}\hbox{\rm Mpc}$. This goes along with a similar mutual alignment amongst voids themselves. 
They have strongly correlated orientations over distances $>30h^{-1}\hbox{Mpc}$ \citep{platen2008}, a scale 
considerably exceeding the typical void size. It forms a strong confirmation of the large scale 
tidal force field as the dominant agent for the evolution and spatial organization of the 
Megaparsec Universe, as emphasized by the Cosmic Web theory \citep{bondweb1996,weybond2008}. 

\begin{figure*}[t]
\begin{center}
\vskip -0.0truecm
\mbox{\hskip -0.75truecm\includegraphics[width=14.25cm]{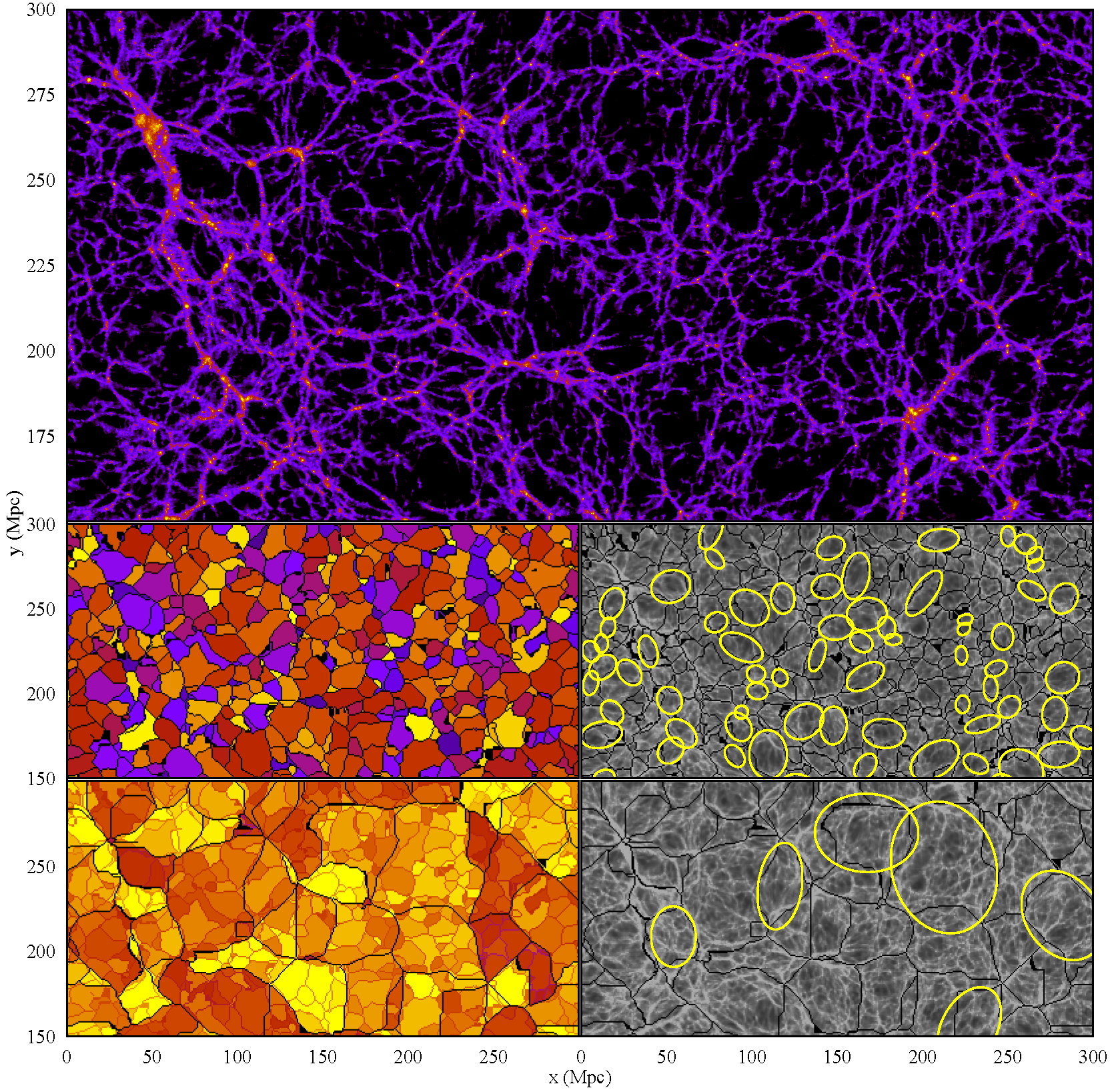}}
\vskip -0.25truecm
\caption{Shapes of voids in a $\Lambda$CDM simulation. Top: a density field slice of thickness $0.25\hmpc$, and size 300 by 150 $\hmpc$. 
\textit{Bottom left two}: the corresponding distribution of voids. The voids have been identified with the Watershed Void Finder 
(Platen et al. 2007), for the top panel a Gaussian filter with radius $R_f = 1.5 \hmpc$, for the bottom panel $R_f = 6.0 \hmpc$ (the 
$1.5\hmpc$ ones are transparently inset). \textit{Bottom right two}: the shapes of the watershed identified voids have been 
determined via the inertia tensor of the corresponding region. A random selection of the ellipsoid fits (yellow) is overlaid 
on the density field (now in grayscale), again at two radii $1.5$ and $6.0\hmpc$.  From Bos et al. 2012.}\label{fig:devoids}
\end{center}
\label{fig:devoids}
\end{figure*}

\subsubsection{Void Shapes and Dark Energy}
\label{sec:devoids}
A third interesting influence on the shape of vois is the possible impact of dark energy on the dynamics and 
evolution of voids. This influence is a result of its direct repulsive effect on the force field of the void, as well 
as indirectly via the external tidal force field induced by the surrounding large scale inhomogeneities. 

Following the earlier suggestions by \cite{parklee2007,leepark2009}, studies by Wandelt and collaborators 
\citep{lavaux2010,biswas2010} showed that void shapes may be used as precision probes of dark energy. 
\cite{biswas2010} even quoted the possibility of improving the figure of the Dark Energy Task Force figure of merit 
by an order of hundred for future experiments like Euclid. An elaborate study by \cite{bos2012} of void shape evolution 
in simulations within a range of different dark energy cosmologies confirmed the high level of sensitivity of 
the shapes of dark matter voids to the underlying dark energy (see fig.~\ref{fig:devoids}). However, it is less 
straightforward to reach similar conclusions of the shapes of voids in the observed galaxy distribution 
\citep{bos2012}. 

\medskip
\cite{lavaux2012} forwarded the suggestion of using the Alcock-Paczysnki test on the stacking of voids 
in a sufficiently large cosmic volume. Assuming that the orientation of voids is random, the shapes of 
properly scaled voids would average out to purely spherical. This comes along with the significant 
advantage of substantially increasing the signal-to-noise of the resulting void stack. Given the 
observation of the voids in a galaxy redshift survey, as a result of the cosmic expansion the resulting void 
stack in redshift space will be stretched differently in the radial direction than in the transverse 
direction. The ratio between the transverse and radial stretching of an intrinsically spherical 
feature yields direct information on the angular diameter distance and Hubble parameter at a redshift z. 
Since the suggestion by \cite{lavaux2012}, several groups realized the large cosmological potential of the 
application of the Alcock-Paczysnki test to voids in large surveys. It has also resulted in a range of 
interesting studies \citep[e.g.][]{sutter2012b,sutter2014,pisani2014a,pisani2014b}, and is generating substantial 
interest for exploiting voids in upcoming large galaxy surveys \citep[e.g.][]{nadathur2016}. 

Of key importance for a succesful application of the Alcock-Paczysnki test to samples of voids is 
to correct for some systematic effects. One concerns the correction for the peculiar velocity outflow 
of the void itself \citep[see][for a clear discussion]{pisani2014b}. Another effect that should be 
accounted for is the alignment of voids on large scales, as discussed in the previous 
subsection~\ref{sec:voidalign}
(also see fig.\ref{fig:devoids}). Clearly, if the region from which the void sample is extracted 
is not much larger than the scale on which tidally induced alignment are expected, the stacked 
voids may not define an intrinsically spherical object and hence influence the outcome of the 
determined cosmological parameters. 

\subsection{Dynamical Influence of Voids}
Various studies have found strong indications for the imprint of voids in the peculiar velocity flows of 
galaxies in the Local Universe. \cite{bothun1992} made the first claim of seeing pushing influence of voids 
when assessing the stronger velocity flows of galaxies along a filament in the first CfA slice. Stronger 
evidence came from the extensive and systematic POTENT analysis of Mark III peculiar galaxy velocities 
\citep{willick1997} in the Local Universe \citep{dekel1990,edbert1990}. POTENT found that for a fully selfconsistent 
reconstruction of the dynamics in the Local Universe, it was inescapable to include the dynamical influence of voids 
\citep{dekel1994}. 

With the arrival of new and considerably improved data samples the dynamical influence of voids in the 
Local Universe has been investigated and understood in greater detail. The reconstruction of the density 
and velocity field in our local cosmos on the basis of the 2MASS redshift survey has 
indeed resulted in a very interesting and complete view of the dynamics on Megaparsec scales.  
This conclusion agree with that reached on the basis of an analysis of the 
peculiar velocity of the Local Group by \cite{tully2008}. Their claim is that the Local Void is 
responsible for a considerable repulsive influence, accounting for $\sim 259$~km~s$^{-1}$ of the 
$\sim 631$~km~s$^{-1}$ Local Group motion with respect to the CMB (also see fig.\ref{fig:localvoids}). 

The substantial dynamical role of voids in the large scale Universe has been most convincingly 
demonstrated in the recent advances in the study of cosmic flows enabled by the completion of 
the Cosmicflows2 and Cosmicflows3 surveys \citep{tully2008,courtois2012}. The implied velocity-based 
reconstructions of the Local Universe, culminating in the suggestion of the hypercluster Laniakea as local 
dynamical entity \citep{tully2014}, clearly reveal the substantial repulsive influence of voids 
in the Local Universe. 

\section{Void Sociology \& Void Hierarchy: \hfill Bubbles in Soapsuds}
\label{sec:voidsocio}
\begin{figure*}
\vskip -0.25truecm
  \begin{center}
     \mbox{\hskip -0.0truecm\includegraphics[width=13.5cm]{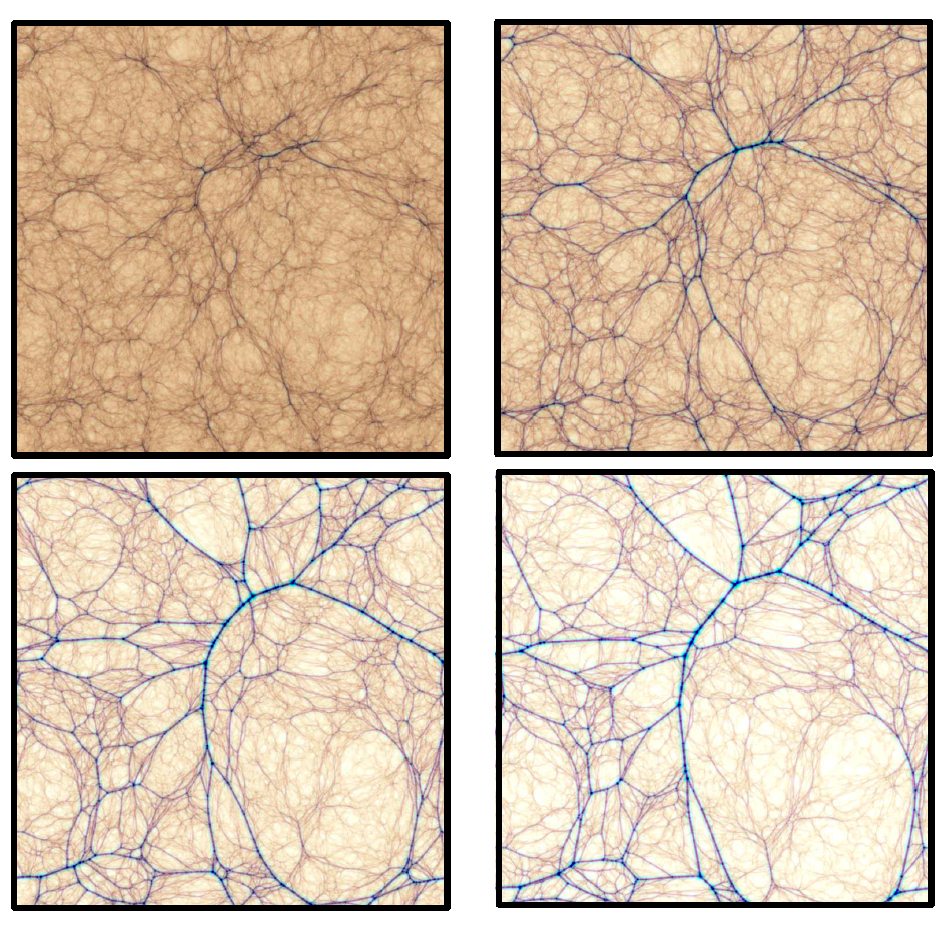}}
\caption{The hierarchical evolution of a void. The evolving mass distribution in and around a large 
void in an LCDM cosmology. The evolution runs from top lefthand to the bottom righthand frame. The 
structure depicted shows the result of an adhesion calculation of the evolving weblike network 
of walls, filaments and nodes in the cosmic volume. It shows in detail and with considerable 
contrast the evolution of the void population. From Hidding et al. 2016.}
\end{center}
\label{fig:voidhieradh}
\end{figure*}

Computer simulations of the gravitational evolution of voids in realistic cosmological environments do 
show a considerably more complex situation than that described by idealized spherical or ellipsoidal models 
\citep{martel1990,regoes1991,dubinski1993,weykamp1993,goldvog2004,colberg2005,padilla2005,ceccar2006,bos2012,aragon2013,
sutter2014b,wojtak2016}. In recent years the huge increase in computational resources has enabled {\rm N}-body 
simulations to resolve in detail the intricate substructure of voids within the context of hierarchical cosmological 
structure formation scenarios \citep{mathis2002,gottloeb2003,goldvog2004,colberg2005,padilla2005,ceccar2006,bos2012,aragon2013,
sutter2014b,wojtak2016}. They confirm the theoretical expectation of voids having a rich substructure as a 
result of their hierarchical buildup (see e.g. fig.~\ref{fig:voidhieradh}). 

\cite{shethwey2004} treated the emergence and evolution of voids within the context of {\it hierarchical} 
gravitational scenarios. It leads to a considerably richer view of the evolution of voids. The role of 
substructure within their interior and the interaction with their surroundings turn out to be essential aspects of 
the {\it hierarchical} evolution of the void population in the Universe. An important guideline are the
heuristic void model simulations by \cite{dubinski1993}, and the theoretical void study by 
\cite{sahni1994} within the context of a Lagrangian adhesion model description. 

In some sense voids have a considerably more complex evolutionary path than overdense halos. Their evolution 
is dictate by two processes: their {\it merging} into ever larger voids as well as the {\it collapse} 
and disappearance of small ones embedded in overdense regions (see fig.~\ref{fig:voidhier}). 
As argued by \cite{shethwey2004}, the implied hierarchical development  of voids, akin to the evolution of overdense halos, 
may be described by an {\it excursion set} formulation \citep{pressschecht1974,bond1991,sheth1998}. To take 
account of the more complex evolutionary history, the evolution of voids needs to be described by a 
{\it two-barrier} excursion set formalism. 

The resulting hierarchical buildup of the void population resembles that of a gradually diluting 
soapsud. At early time, emerging from a primordial Gaussian random field, the first bubbles to 
appear are small matured voids. The ones coexisting within the realm of a larger underdense region, will 
gradually merge into a larger void. This is the typical fate of a void in a {\em void-in-void} configuration. 
Meanwhile, small voids that find themselves in overdense regions or near the boundary of larger 
underdensities get squeezed out of existence, the fate for a void in a {\it void-in-cloud} configuration. 
What remains is a sud of larger void bubbles. This sequence proceeds continuously. Given the almost 
volume-filling nature of voids, this void hierarchical process conjures up the impression of a weblike 
pattern evolving as the result of continuously merging bubbles, with filaments, walls and cluster nodes 
at the interstices of the network \citep[see][for a geometric model along these lines]{ickewey1987,aragon2014}. 

\subsubsection{Void Merging}
First, consider a small region which was less dense than the critical void density value.  It may be that 
this region is embedded in a significantly larger underdense region which is also less dense than the critical density. 
Many small primordial density troughs may exist within the larger void region. Once small density depressions located 
within a larger embedding underdensity have emerged as true voids at some earlier epoch, their expansion tends 
to slow down. 

\begin{figure*}[t]
\begin{center}
\vskip -0.0truecm
\mbox{\hskip -0.0truecm\includegraphics[width=13.5cm]{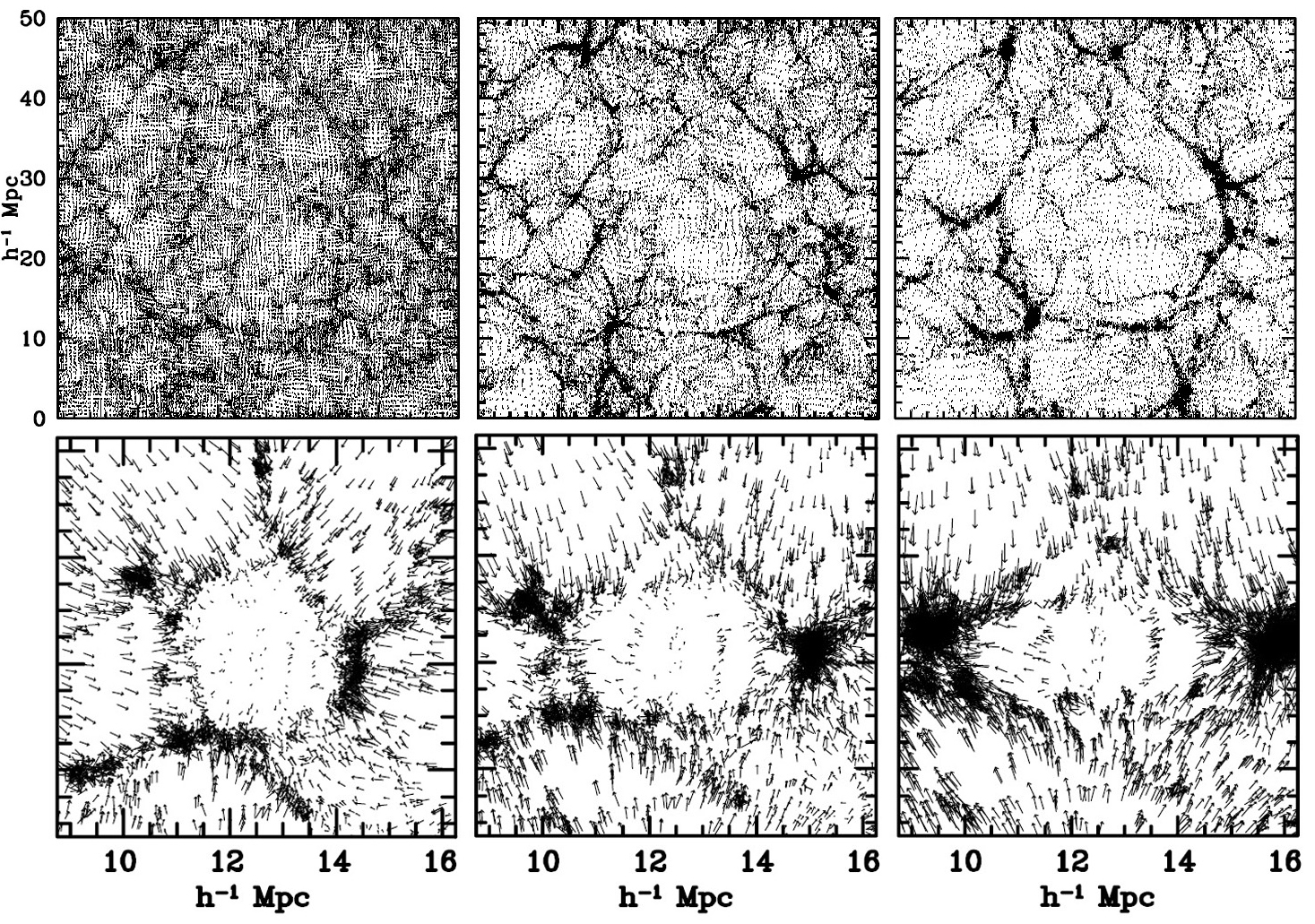}}
\vskip 0.0truecm
\caption{The two modes of void evolution: void merging (top row) and void collapse (bottom row). 
Top: three timesteps of evolving void structure in a 128$^3$ particle {\rm N}-body simulation of structure 
formation in an SCDM model ($a_{\rm exp}=0.1, 0.3,0.5$). The sequence shows the gradual development 
of a large void of diameter $\approx 25h^{-1}\hbox{Mpc}$ as the complex pattern of smaller voids and 
structures which had emerged within it at an earlier time, merge with one another. It illustrates 
the {\it void-in-void} process of the evolving void hierarchy. Bottom: a choice of three collapsing 
voids in a constrained {\rm N}-body simulation, each embedded within an environment of different tidal 
shear strength. The arrows indicate the velocity vectors, showing the infall of outer regions onto 
the void region. As a result the voids will be crushed as the surrounding matter rains down on them.}
\label{fig:voidhier}
\end{center}
\end{figure*}

When the adjacent subvoids meet up, the matter in between is squeezed in thin walls and filaments. The peculiar 
velocities perpendicular to the void walls are mostly suppressed so that the flow of matter is mostly 
confined to tangential motions. The subsequent merging of the subvoids is marked by the gradual fading of 
these structures while matter evacuates along the walls and filaments towards the enclosing boundary of 
the emerging void \citep{dubinski1993} (also see top row fig.~\ref{fig:voidhier}). The timescale on which 
the internal substructure of the encompassing void is erased is approximately the same as that on which it 
reaches maturity.

The final result is the merging and absorption of the subvoids in the larger void emerging from the embedding 
underdensity. Hence, as far as the void population is concerned only the large void counts, while the 
smaller subvoids should be discarded as such. Only a faint and gradually fading imprint of their original outline 
remains as a reminder of the initial internal substructure. 

\subsubsection{Void Collapse} 
A {\it second} void process is responsible for the radical dissimilarity between void and halo populations. If a small 
scale minimum is embedded in a sufficiently high large scale density maximum, then the collapse of the larger surrounding 
region will eventually squeeze the underdense region it surrounds: the small-scale void will vanish when the region around 
it has collapsed completely. Alternatively, though usually coupled, they may collapse as a result of the 
tidal force field in which they find themselves. If the void within the contracting overdensity has been 
squeezed to vanishingly small size it should no longer be counted as a void (see fig.~\ref{fig:voidhier}, bottom row). 

When inspecting the evolving matter distribution in and around voids, such as in the high resolution configuration 
of fig.~\ref{fig:voidhieradh}, we find that most of the collapsing or squeezed voids are small voids to be found in and near the 
boundary regions of large underdense void regions. The small voids in these regions are strained by the high density 
boundary regions, mostly filaments and clusters, and in many situations squeezed out of existence. On the other hand, 
subvoids in the interior of a larger void tend to merge with surrounding peers. A detailed 
assessment of the evolution of these regions reveals that the void collapse process is an important aspect of 
the evolution and buildup of the cosmic web. Interestingly, recent observational studies have indeed identified 
the existence of collapsing void regions by studying the redshift structure around voids in the SDSS galaxy 
redshift survey \citep{paz2013}. 

The collapse of small voids is an important aspect of the symmetry breaking between underdensities and 
overdensities. In the primorial Universe, Gaussian primordial conditions involve a perfect symmetry between under- and 
overdense. Any inspection of a galaxy redshift map or an {\rm N}-body simulation shows that there is a marked difference 
between matter clumps and voids. While the number density of halos is dominated by small objects, void collapse 
is responsible for the lack of small voids.

\subsection{Void Excursions}
\label{sec:voidexcur}
Marked by the two processes of merging and collapse, the complex hierarchical buildup of the 
void population may be modelled by a two-barrier excursion set formalism \citep{shethwey2004}. 
The barriers refer to the critical (linear) density thresholds involved with the {\it merging} 
and {\it collapse} of voids. Whenever the linearly extrapolated (primordial) $\delta_L({\vec r},t|R)$ 
on a scale $R$, 
\begin{equation}
\delta_L(r,t|R)\,=\,{\displaystyle D(t) \over \displaystyle D(t_i)}\,\delta_L(r,t_i|R)\,,
\end{equation}
exceeds a density threshold $f_c$ it will collapse. For an Einstein-de Sitter $\Omega_m=1$ Universe the 
critical value has the well-known value $f_c\,\simeq\,1.686$ \citep{gunngott1972}. A void will form 
when an underdensity reaches the critical density threshold of {\it shell-crossing}, corresponding to a 
value of $f_v\,\simeq\,-2.81$ for spherical voids in an Einstein-de Sitter Universe. The linear density growth factor 
$D(t)$, normalized to unity at the present epoch, follows from the integral \citep{heath1977,peebles1980,hamilton2001,
lahavsuto2004},
\begin{eqnarray}
D(t)\,=\,D(t,\Omega_{m,0},\Omega_{\Lambda,0})\,=\,{\displaystyle 5\,\Omega_{m,0} H_0^2 \over \displaystyle 2}\,H(a)\,\int_0^a\,{\displaystyle da' \over \displaystyle a'^3 H^3(a')}\,.
\label{eq:lingrowthlambd}
\end{eqnarray}

\noindent Emerging from a primordial Gaussian random field, many small voids may coexist within one larger void.
Small voids from an early epoch will merge with one another to form a larger void at a later epoch. The excursion set 
formalism takes account of this {\em void-in-void} configuration by discarding these small voids from the void count. 
To account for the impact of voids disappearing when embedded in collapsing regions, the two-barrier formalism also deals with 
the {\it void-in-cloud} problem. 

By contrast, the evolution of overdensities is governed only by the {\it cloud-in-cloud} process: the {\it cloud-in-void} process is 
much less important, because clouds which condense in a large scale void are not torn apart as their parent void expands around them. 
This asymmetry between how the surrounding environment affects halo and void formation is incorporated into the {\it excursion set approach} 
by using one barrier to model halo formation and a second barrier to model void formation. Only the first barrier matters for halo formation, 
but both barriers play a role in determining the expected abundance of voids. 

\subsection{Void Population Statistics.}
\begin{figure*}
\vskip -0.25truecm
  \begin{center}
     \mbox{\hskip -0.0truecm\includegraphics[width=13.5cm]{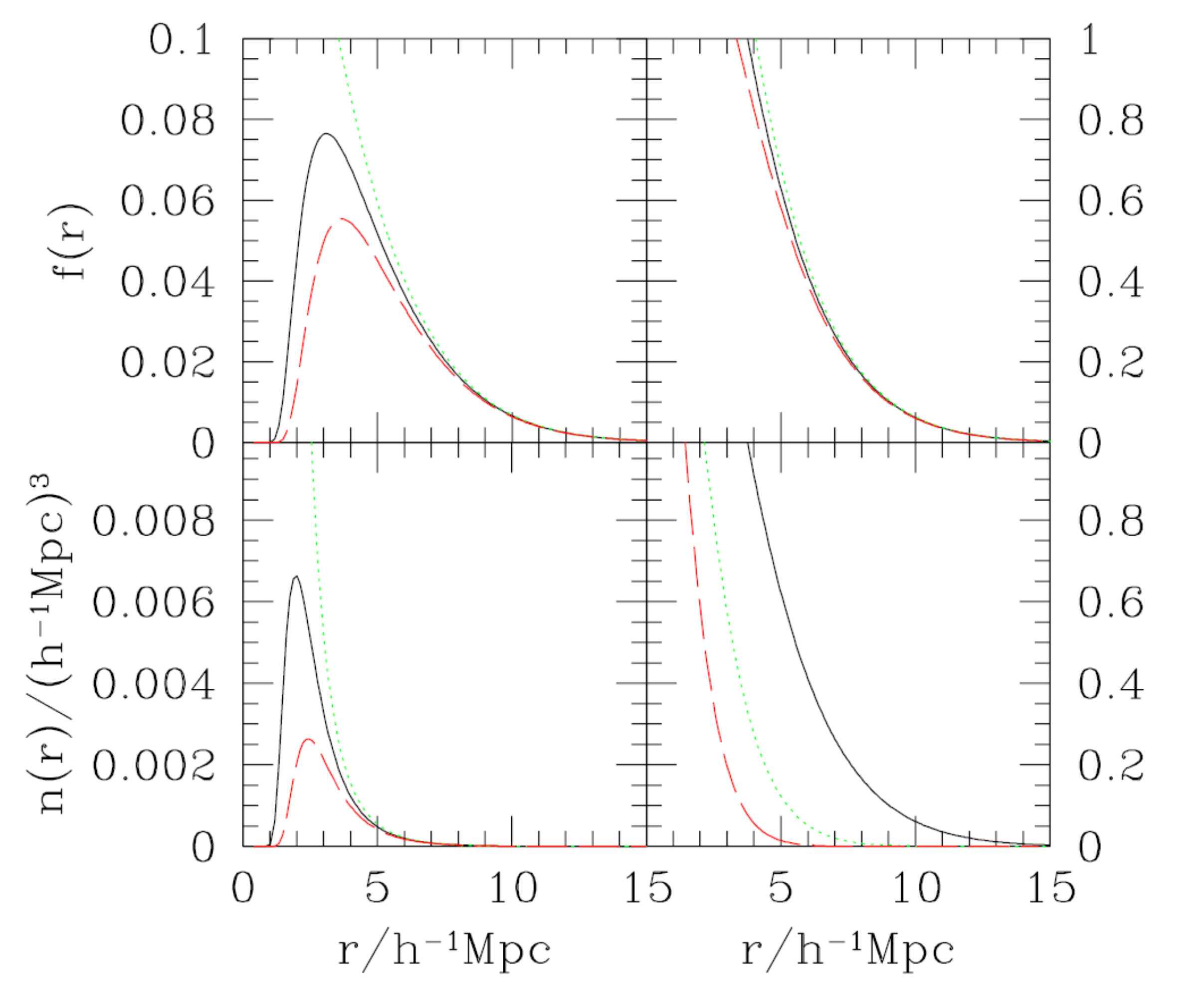}}
\caption{Distribution of void radii predicted by the two-barrier 
  extended PS formalism, in an Einstein de-Sitter model with 
 $P(k)\propto k^{-1.5}$, normalized to $\sigma_8=0.9$ at $z=0$.  
 Top left panel shows the mass fraction in voids of radius $r$.  
 Bottom left panel shows the number density of voids of radius $r$.  
 Note that the void-size distribution is well peaked about a 
 characteristic size provided one accounts for the void-in-cloud 
 process.  Top right panel shows the cumulative distribution of the 
 void volume fraction. 
 Dashed and solid curves in the top panels and bottom left panel 
 show the two natural choices for the importance of the void-in-cloud 
 process discussed in the text:  $\delta_{\rm c}=1.06$ and 1.686, 
 with $\delta_{\rm v}=-2.81$.  Dotted curve shows the result of 
 ignoring the {\em void-in-cloud} process entirely.
 Clearly, the number of small voids decreases as the ratio of 
 $\delta_{\rm c}/|\delta_{\rm v}|$ decreases. 
 Bottom right panel shows the evolution of the cumulative void volume 
 fraction distribution. The three curves in this panel are for 
 $\delta_{\rm c}=1.686(1+z)$, where $z=0$ (solid), 0.5 (dotted) 
 and~1 (dashed). From Sheth \& van de Weygaert 2004.}
\end{center}
\label{fig:voidhiersvdw}
\end{figure*}

The analytical evaluation of the two-barrier random walk problem in the extended Press-Schechter approach leads 
directly to a prediction of the distribution function $n_v(M)$ for voids on a mass scale $M$ (or corresponding void 
size $R$\footnote{The conversion of the void mass scale to equivalent void radius $R$ is done by assuming the simplest 
approximation, that of the spherical tophat model. According to this model a void has expanded by a factor of 1.7 by 
the time it has mature, so that $V_v=(M/\rho_u)*1.7^3$.}). The resulting void spectrum\footnote{Note that for 
near-empty voids the mass scale $M$ is nearly equal to the corresponding void mass deficit.}. is peaked, with a sharp 
cutoff at both small and large values of the peak mass $M_{v,\ast}$ (fig.~\ref{fig:voidhiersvdw}, bottom lefthand frame),
\begin{eqnarray}
n_v(M)\,{\rm d}M&&\,\approx\nonumber\\
\ \\
&&\sqrt{\frac{2}{\pi}}\,\frac{\rho_u}{M^2}\,\nu_v(M)\,\exp\left(-\frac{\nu_v(M)^2}{2}\right)\,
\left|\frac{{\rm d} \ln \sigma(M)}{{\rm d} \ln M}\right|\,\exp\left\{-{|f_{\rm v}|\over f_{\rm c}}\,
 {{\cal D}^2\over 4\nu_v^2}-2{{\cal D}^4\over\nu_v^4} \right\}\,,\nonumber
\end{eqnarray}
where $\sqrt{\nu_v(M)}$ is the fractional relative underdensity, 
\begin{equation}
\nu_v(M)\,\equiv\,\frac{{|f_{\rm v}|}}{\sigma(M)}\,,
\label{eq:numvoid}
\end{equation}
\noindent in which with the dependence on the mass scale $M$ entering via the rms density fluctuation on that scale, 
$\sigma(M)$. The quantity ${\cal D}$ is the ``{\it void-and-cloud parameter}'', ${\cal D} \equiv |f_{\rm v}|/(f_{\rm c}+|f_{\rm v}|)$.  
It parameterizes the impact of halo evolution on the evolving population of voids: the likelihood of smaller 
voids being crushed through the {\it void-in-cloud} process decreases as the relative value of the collapse barrier 
$f_{\rm c}$ with respect to the void barrier $f_{\rm v}$ becomes larger. 

Assessment of the relation above shows that the population of large voids is insensitive to the {\it void-in-cloud} 
process. The large mass cutoff of the void spectrum is similar to the ones for clusters and reflects the Gaussian nature of the 
fluctuation field from which the objects have condensed. The {\em characteristic void size} increases with time: the gradual 
merging of voids into ever larger ones is embodied in a self-similar shift of the peak of the void spectrum.

When evaluating the corresponding fraction of contained in voids on mass scale M, $f(M)\,=\,{M\,n_v(M)/\rho_u}\,$, we 
find that this also peaks at the characteristic void scale. It implies a mass fraction in voids of approximately thirty 
percent of the mass in the Universe, with most of the void mass to be found in voids of this characteristic mass. 

While the two-barrier excursion set formalism offers an attractive theoretical explanation for the 
distinct asymmetry between clumps and voids and for the peaked void size distribution, realistic 
cosmological simulations are needed to identify where the disappearing small-scale voids are to be found 
in a genuine evolving cosmic matter distribution. 

An important aspect of the implied void population is that it is approximately {\it space-filling}. It underlines 
the adagio that the large scale distribution of matter may be compared to a {\it soapsud of expanding bubbles}. 
This follows from evaluation of the cumulative integral 
\begin{equation}
{\cal F}_V(M)\,\equiv\,\int_M^{\infty}\,(1.7)^3\,{\frac{M'\,n_v(M')}{\rho_u}}\,dM'\,\,.   
\end{equation}
where the factor 1.7 is an estimate of the excess expansion of the void based upon the spherical model for void 
evolution (see footnote). The top righthand panel of fig.~\ref{fig:voidhiersvdw} shows the resulting (current) 
cumulative void volume distribution: for a finite value of void radius $R$ the whole of space indeed appears to be occupied by voids, 
while there is the gradual shift of the cumulative volume distribution towards larger voids. 
In other words, the correct image appears to be that of a gradually unfolding bubbly universe in which the 
average size of the voids grows as small voids merge into ever larger ones. 

\medskip
While the SvdWmodel has proven to provide a good model for the evolving void 
population, it should be considered as a basic framework. Substantial finetuning to realistic 
circumstances, devoid of spherical symmetry and other implicit simplifications, will likely 
improve the model. One major issue concerns the value of the typical void formation threshold or 
density barrier. The simple choice for the shellcrossing of a hypothetical spherical configuration 
certainly deserves scrutiny, and considerable improvements to this have been forwarded 
\citep{paranjape2012,jennings2013,pontzen2016}. 

Also, to connect the SvdW formalism to the voids in the observed galaxy distribution involves 
a range of factors. Galaxies are diluted and biased tracers of the underlying dark matter 
distribution. While the dark matter voids define a rich and complex structural pattern 
\citep[see e.g.][]{leclercq2015,leclercq2015b}, voids in the galaxy distribution will only represent a 
limited aspect of the void populations. \cite{furlpir2006} elaborated on the SvdW formalism 
to describe what it would imply for voids in the galaxy distribution.

The advantage of having an analytical formalism for describing the void population has recently 
been recognized in the context of casting constraints on the nature of dark energy. By evaluating 
the implied void population in a range of dark energy scenarios, \cite{pisani2015} demonstrated that 
in upcoming surveys such as Euclid it will be possible to infer competitive estimates of the 
equation of state of dark energy, as well as on its time derivative. The use of the analytical 
description of the two-barrier excursion set model provides highly versatile and flexible 
framework for readily identifying the most sensitive factors and situations. The same philosophy 
was followed in the elaboration by \cite{li2012}, who applied the excursion set formalism 
towards a general description of the evolving void population in modified gravity scenarios.

\subsection{Void Substructure}
An important issue within the hierarchically proceeding evolution of voids and the Cosmic Web 
is the fate of its substructure. In voids the diluted and diminished infrastructure remains 
visible, at ever decreasing density contrast, as cinders of the earlier phases of the {\it void 
hierarchy} in which the substructure stood out more prominently.

{\rm N}-body simulations show that voids do retain a rich infrastructure. Examples such as the 
fig.~\ref{fig:voidhieradh}, and images of large cosmological simulations such as the Millennium 
and the Illustris simulation \citep{springmillen2005} show that while void substructure does fade, 
it does not disappear. We may find structures ranging from 
filamentary and sheetlike structures to a population of low mass dark matter halos and galaxies 
\cite[see e.g.][]{rieder2013}. Although challenging, these may also be 
seen in the observational reality. The galaxies that populate the voids do currently attract quite some 
attention (see next section). Also, the SDSS galaxy survey has uncovered a substantial level of 
substructure within voids, such as a the VGS31 filament by \cite{beygu2013}.

The gradual fading of the initially rich substructure as a void expands, and its matter 
Rcontent evacuates its interior, is clearly visible in the development of the void region in 
fig.~\ref{fig:lcdmvoid}. This may be seen as the slowing of structure growth in the 
low-density environment of voids. The resulting observation is one in which the density distribution in a void region 
undergoes a manifest and continuous transformation towards an ever larger dominating scale, 
while it mostly retains the topological features of the initial density field. 
In a sense, it reflects of the buildup of the Cosmic Web itself, its basic pattern 
already recognizable in the primordial matter distribution \citep{bondweb1996,weybond2008}.
\acknowledgements
I wish to thank the many collaborators and students that over many years have accompanied 
me in my explorations of the desolate but wildly interesting void regions in our universe. 
In particular I wish to thank Ravi Sheth, Johan Hidding, Marius Cautun, Patrick Bos, Bernard Jones, 
Erwin Platen, Mathijs Dries, Francisco Kitaura and Vincent Icke for major contributions to the work 
described in this paper. RvdW is particularly grateful to J. Hidding and F. Kitaura for the use 
of the images in figures 3 and 4 in advance of publication and to J. Hidding for figure 11 
in advance of publication.

\end{document}